\documentclass[lettersize,journal]{IEEEtran}
\usepackage{lipsum}

\usepackage{framed}
\IEEEoverridecommandlockouts
\usepackage{cite}
\usepackage{url}
\usepackage[htt]{hyphenat}
\usepackage{multicol}
\usepackage{tcolorbox}
\usepackage{subcaption}
\usepackage{amsmath,amssymb,amsfonts}
\usepackage{graphicx}
\usepackage{subcaption}
\usepackage{textcomp}
\usepackage[normalem]{ulem} 
\usepackage{xcolor}
\usepackage{listings}
\usepackage{multirow}
\usepackage[linesnumbered,ruled,vlined]{algorithm2e}
\usepackage{xcolor}
\usepackage{caption}
\usepackage{setspace}
\usepackage{array}
\usepackage{quoting}
\usepackage{mdframed}
\usepackage{syntax}
\usepackage{hyperref}
\usepackage{diagbox}
\usepackage{multirow}
\usepackage{CJK}
\usepackage{mathrsfs}
\def\BibTeX{{\rm B\kern-.05em{\sc i\kern-.025em b}\kern-.08em
    T\kern-.1667em\lower.7ex\hbox{E}\kern-.125emX}}
\usepackage{mathdots}
\usepackage{autobreak}
\usepackage{booktabs}
\usepackage{balance}
\newcommand{\tabincell}[2]{\begin{tabular}{@{}#1@{}}#2\end{tabular}}
\usepackage{breqn}
\usepackage{bbding}
\definecolor{commentgreen}{RGB}{2,112,10}
\definecolor{eminence}{RGB}{108,48,130}
\definecolor{weborange}{RGB}{255,165,0}
\definecolor{frenchplum}{RGB}{129,20,83}
\usepackage{comment}

\begin{document}

\lstset{
language=SQL,
 tabsize=2,
 basicstyle=\tt\scriptsize,
 columns=fixed,       
 numbers=left,
 numbersep=3pt,
 numberstyle=\tiny\color{darkgray},
 frame=none,
 backgroundcolor=\color[RGB]{245,245,245},
 keywordstyle=\bfseries,
 commentstyle=\it\color[RGB]{0,96,96},
 stringstyle=\rmfamily\slshape\color[RGB]{128,0,0},
 showstringspaces=false,
 language=SQL,
 morekeywords={FUNCTION, DECLARE, IF, ELSIF, RETURN, FOR, LOOP, LANGUAGE, SQL, Prolog, Facts, Rules, Queries},
 escapechar=@,
 breaklines=true,
 breakatwhitespace=true,
}

\newcommand{\tool}{\textsc{SemConT}}
\newcommand{\ls}[1]{\textcolor{orange}{LS: #1}}

\newcommand{\tcl}[1]{\textcolor{blue}{TCL: #1}}

\title{Conformance Testing of Relational DBMS Against SQL Specifications}


\author{\IEEEauthorblockN{1\textsuperscript{st} Shuang Liu}\\
\IEEEauthorblockA{\textit{ School of Information,}
\textit{ Renmin University of China,}\\
Beijing, China. \\
 E-mail: shuang.liu@ruc.edu.cn}
\\
~\\
\and
\IEEEauthorblockN{2\textsuperscript{nd} Chenglin Tian}\\
\IEEEauthorblockA{\textit{ School of Computer Science,}
\textit{ Beijing University of Posts and Telecommunications,}\\
Beijing, China. \\
 E-mail: 2024010271@bupt.cn}
\\
~\\
\and
\IEEEauthorblockN{3\textsuperscript{rd} Jun Sun}\\
\IEEEauthorblockA{\textit{Singapore Management University,} \\
Singapore, Singapore. \\
 E-mail: junsun@smu.edu.sg}
 \\
 ~\\
\and
\IEEEauthorblockN{4\textsuperscript{th} Ruifeng Wang}\\
\IEEEauthorblockA{\textit{ the College of Intelligence and Computing,}
\textit{ Tianjin
University,}\\
Tianjin, China. \\
 E-mail: ruifeng_wong@tju.edu.cn}
 \\
 ~\\
\and
\IEEEauthorblockN{5\textsuperscript{th} Wei Lu}\\
\IEEEauthorblockA{\textit{ School of Information,} 
\textit{ Renmin University of China,}\\
Beijing, China. \\
 E-mail: lu-wei@ruc.edu.cn}
 \\
 ~\\
\and
\IEEEauthorblockN{6\textsuperscript{th} Yongxin Zhao}\\
\IEEEauthorblockA{\textit{East China Normal University, } \\
Shanghai, China. \\
 E-mail: yxzhao@sei.ecnu.edu.cn}
  \\
  ~\\
\and
\IEEEauthorblockN{7\textsuperscript{th} Yinxing Xue}\\
\IEEEauthorblockA{\textit{University of Science and Technology of China, } \\
Suzhou, China. \\
 E-mail: yxxue@ustc.edu.cn}
 \\
 ~\\
\and
\IEEEauthorblockN{8\textsuperscript{th} Junjie Wang}\\
\IEEEauthorblockA{\textit{ the College of Intelligence and Computing,} 
\textit{ Tianjin
University,}\\
Tianjin, China. \\
 E-mail: junjie.wang@tju.edu.cn}
 \\
 ~\\
\and
\IEEEauthorblockN{9\textsuperscript{th} Xiaoyong Du}\\
\IEEEauthorblockA{\textit{ School of Information,} 
\textit{ Renmin University of China,}\\
Beijing, China. \\
 E-mail: duyong@ruc.edu.cn}
}

\maketitle

\begin{abstract}

A Relational Database Management System (RDBMS) is one of the fundamental software that supports a wide range of applications, making it critical to identify bugs within these systems.  
There has been active research on testing RDBMS, most of which employ crash or use metamorphic relations as the oracle.
Although existing approaches can detect bugs in RDBMS, they are far from comprehensively evaluating the RDBMS's correctness (i.e., with respect to the semantics of SQL).
In this work, we propose a method to test the semantic conformance of RDBMS i.e., whether its behavior respects the intended semantics of SQL. 
Specifically, we have formally defined the semantics of SQL and implemented them in Prolog.
Then, the Prolog implementation serves as the reference RDBMS, enabling differential testing on existing RDBMS. 
We applied our approach to four widely-used and thoroughly tested RDBMSs, i.e., MySQL, TiDB, SQLite, and DuckDB. 
In total, our approach uncovered 19 bugs and 13 inconsistencies, which are all related to violating the SQL specification or missing/unclear specification, thereby demonstrating the effectiveness and applicability of our approach. 
\end{abstract}
\section{Introduction}
\label{sec:introduction}

Relational Database Management System (RDBMS)~\cite{mysql,tidb,sqlite,psql,duckdb} are widely adopted in various applications, including both web applications and embedded systems~\cite{amazon,ebay,booking,facebook}.
Structured Query Language (SQL) is the standard programming language for relational databases, and its specification is formally documented in ISO/IEC 9075:2016~\cite{iso-iec-9075-2016}.
As the parser and executor of SQL queries, an RDBMS should conform to the SQL specification to ensure the correct implementation of the semantics.
As of 2023, there are more than 416 different implementations of relational databases, yet many of these implementations deviate from the specification~\cite{slutz1998massive,zhu2022fuzzing,ceri1985translating}.
Bugs have also been reported due to violations of the SQL specifications~\cite{3f085531bf}, which may potentially lead to data integrity issues or even security vulnerabilities in the database.  
The serious impacts of bugs in RDBMS have been discussed by various existing studies~\cite{rigger2020detecting, rigger2020finding, sqlancer, liang2022detecting}, and thus it is critical to detect those bugs. 

\begin{figure}[t]
\centering
\begin{subfigure}{0.45\textwidth}
\begin{lstlisting}
SELECT 3>(-5|-4); 
   --expected: 1 @\color{green}{\CheckmarkBold}@, actual: 0 @\color{red}{\XSolidBrush}@
\end{lstlisting}
\caption{A test case triggering bug (ID 39259) in TiDB 6.6.0 }
\label{fig:motivation-a}
\end{subfigure}
\begin{subfigure}{0.45\textwidth}
\begin{lstlisting}
SELECT 'Hello'||NULL;
   --SQLite: NULL, PostgreSQL: 'Hello'
\end{lstlisting}
\centering
\includegraphics[width=\textwidth,height=0.15\textwidth]{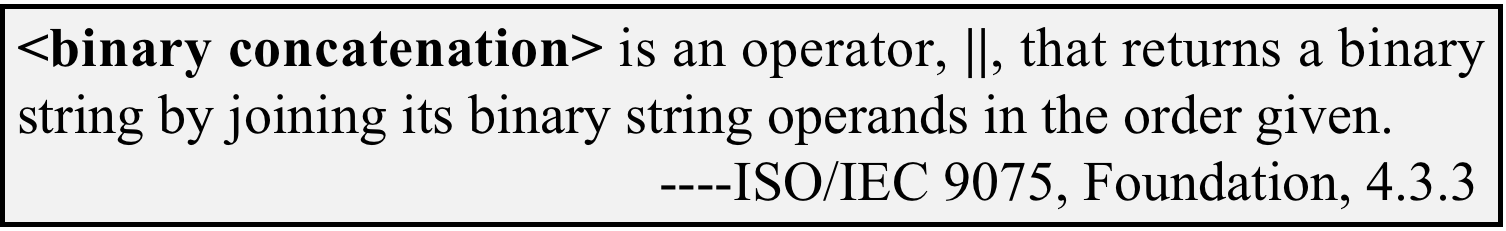}
\caption{A test case causing an inconsistency between SQLite version 3.39.0 and PostgreSQL version 16.2, and the SQL specification  
}
\label{fig:motivation-b}
\end{subfigure}
\caption{A Bug and an Inconsistency detected by our approach}
\label{fig:motivation}
\end{figure}

As the golden standard for correct SQL behavior, the specification of SQL should be clearly described, and an RDBMS should conform to the SQL specification. 
Inconsistencies between RDBMS implementations and the specification can lead to unexpected results.
Figure~\ref{fig:motivation} presents two motivating examples, one bug and one inconsistency that our approach detected.
Figure~\ref{fig:motivation-a} is a SQL query that triggers a bug in TiDB version 6.6.0\footnote{We have also detected this bug in MariaDB 10.9.4 and MySQL 8.0.29. }, which occurs when performing a bitwise operation on negative numbers, which are by default signed numbers. 
The root cause is that TiDB incorrectly represents the result of bitwise OR (\texttt{|}) on two signed 64-bit integer (\texttt{-5} and \texttt{-4}) as an unsigned 64-bit integer (18446744073709551615). 
Regarding this bug, the expected result of \texttt{-5|-4} is the signed integer \texttt{-1}, the binary representation of which is 64 bits of 1. 
However, as TiDB treats the result of bitwise OR as an unsigned integer, it returns the unsigned 64-bit integer 18446744073709551615, which is the decimal representation of the binary number of 64 bits 1. 
Therefore, \texttt{3>(-5|-4)} is evaluated to 0 (false) in TiDB.
For other RDBMSs that we have tested, e.g., SQLite and PostgreSQL, the given query is correctly executed (\texttt{-5|-4} returns \texttt{-1} and \texttt{3>(-5|-4)} returns \texttt{1}).
The underlying reason for this inconsistency is that the bitwise operation on signed numbers in the SQL specification is under specified. Therefore, different RDBMSs have different implementations. 
In practice, large online systems, such as that of Alibaba and Tencent, may integrate and adopt different RBDMSs (sometimes dynamically) for efficiency reasons, and such bugs may result in unexpected system behaviors. 
To avoid problems caused by such inconsistencies between RDBMS implementations, we need a method of systematically identifying such under-specification in SQL semantics.

Figure~\ref{fig:motivation-b} shows an inconsistency between two RDBMS implementations, SQLite and PostgreSQL, when dealing with a \texttt{NULL} value.
In SQLite, concatenating any string with \texttt{NULL} returns \texttt{NULL} by default. 
In contrast, PostgreSQL treats \texttt{NULL} as an empty string.
Thus, concatenating it with string \texttt{Hello} results in \texttt{Hello}. 
In the SQL specification, there in only one sentence describing the binary concatenation operator (as shown in Figure~\ref{fig:motivation-b}), and it fails to clearly specify how to process \texttt{NULL}, which is a specific data type in SQL representing unknown data. 
This inconsistency poses substantial challenges for users of database systems. 
When transitioning between databases and leveraging features with inconsistent implementations, users may encounter unexpected outcomes. 


%

%

From the motivational examples, we observe that failing to respect the SQL specification or unclearly documented   specifications can result in bugs or inconsistent implementations across different RDBMSs, potentially confusing users.  
Therefore, it is critical to test the conformance of RDBMS implementations with the SQL specification. 
However, existing approaches on testing RDBMSs either use different RDBMSs as the test oracle~\cite{slutz1998massive} or propose metamorphic relations~\cite{liang2022detecting, rigger2020testing, rigger2020detecting, rigger2020finding}. 
None of those approaches consider testing the conformance of RDBMS implementations with SQL specifications. 
Consequently, they are only scratching the surface in evaluating the correctness of RDBMS.



To address the issue of automatic conformance testing between SQL specifications and RDBMS implementations, two main challenges arise. 
Firstly, the SQL specification is written in natural language, which is not directly executable. 
Secondly, it is challenging to generate test queries that comprehensively cover all aspects defined in the SQL specification, including descriptions of keywords and parameters. 
This complexity makes comprehensive conformance testing a challenging task.


In this work, we propose the first automatic conformance testing approach for relational DBMSs. 
To address the first challenge, we develop a formal denotational semantics of SQL and implemented the  formalized semantics in Prolog. 
This executable SQL semantics is then used as an oracle to detect inconsistencies between RDBMS implementations and the SQL specification. 
To overcome the second challenge, we propose three coverage criteria based on the defined semantics, which are then utilized to guide the test query generation process, ensuring comprehensive coverage of SQL specifications.


To evaluate the effectiveness of our approach, we conducted experiments on four popular and well-tested RDBMSs,  successfully detecting 19 bugs—18 of which are reported for the first time—and 13 inconsistencies.   
Further examination revealed that 8 bugs and 2 inconsistencies are due to deviations in RDBMS implementations from the SQL specification, 11 bugs and 11 inconsistencies are attributed to missing or unclear descriptions in the SQL specification itself. 
Additionally, we evaluated the effectiveness of the proposed coverage criteria. 
The results indicate that all three coverage criteria contribute to generating more diverse test queries, which in turn help uncover more bugs and inconsistencies, and combining all three coverage criteria yields the most effective testing results.   

To summarize, in this work, we make the following contributions. 
\begin{itemize}
    \item We propose the first method, \tool{}, on semantic conformance testing of RDBMS implementations with the SQL specification, for which we formalize the semantics of SQL and implement the formalized semantics in Prolog, enabling automatic conformance testing.
    \item We introduce three coverage criteria based on the formalized semantics, which effectively guide test query generation.
    \item We evaluate \tool{} on four popular and thoroughly-tested RDBMSs, and detected 19 bugs, 18 of which are reported for the first time, and 13 inconsistencies. 
    \item We released our Prolog implementation at \url{https://github.com/DBMSTesting/sql-prolog-implement} to inspire further research in this area. 
\end{itemize} 
\section{PRELIMINARY}
\label{sec:preliminary}

\subsection{Prolog}

Prolog (Programming in logic)~\cite{clocksin2003programming} is a logical programming language based on first-order predicate calculus that focuses on deductive reasoning. 
%
A Prolog program consists of three components, i.e., facts, rules, and queries.
Facts and rules describe the axioms of a given domain, while queries represent propositions to be proven. 
In the context of data and relationships, facts and rules define the logic and relations of a given domain. 
Computations are then conducted by applying queries to these facts and rules. 
Similarly, when facts and rules are used to capture the laws governing state changes, queries represent the desired target state. 


\begin{figure}[t]
\begin{lstlisting}
Facts:
Tables=[[t,[a,b],[1,4],[2,5],[3,8]]]
Rules:
select_clause((@\normalcolor{null}@),Tb,[]).
select_clause(X,Tb,Z) :-
    isConstant(X),
    add_X(X,Tb,T),
    column_select(X,T,Z).
select_clause(X,Tb,Z) :-
    list(X),
    column_select(X,Tb,Z).
...
from_clause(T,Z) :-
    list(T),
    table_select(T,Tables,Z).   
...
Queries:
from_clause(t,TableList).
select_clause(t.b,TableList,Filtered).
   --Return Result: Filtered = [[4],[5],[8]].
\end{lstlisting}
\captionsetup{justification=centering}
\caption{An example of implementing the semantics of SQL keywords \texttt{SELECT} and \texttt{FROM} using Prolog}
\label{fig:prolog-demo}
\end{figure} 

\begin{figure*}[htbp]
\begin{framed}
\leftline{$
(1)\langle queryexp\rangle       ::= \{\langle collection\ clause\rangle\ |\ \langle select\ clause\rangle       \langle from\ clause\rangle  [ \langle where\ clause\rangle       ][ \langle group\ by\ clause\rangle       ]     
$}

\leftline{$
\ \ \ \ \ [ \langle having\ clause\rangle       ]\} [ \langle order\ by\ clause\rangle       ]
$}

\leftline{$(2)\langle collection\ clause\rangle       ::=  \langle queryexp\rangle       \langle cop\rangle       \langle queryexp\rangle       $}
\leftline{$(3)\langle cop\rangle       ::= \text{UNION}\ [ \text{ALL} ]| \text{EXCEPT}\ [ \text{ALL} ]|\text{INTERSECT}\ [ \text{ALL}]$}

\leftline{$(4)\langle from\ clause\rangle       ::= \text{FROM} \langle tref\rangle       [ { , \langle tref\rangle       }... ]  $}
\leftline{$(5)\langle tref\rangle       ::= \langle tname\rangle\ |\ \langle joined\ table\rangle         $}
\leftline{$(6)\langle joined\ table\rangle       ::= \langle cross\ join\rangle\ |\ \langle qualified\ join\rangle\ |\ \langle natural\ join\rangle       $}
\leftline{$(7)\langle cross\ join\rangle       ::= \langle tname\rangle       \text{CROSS}\ \text{JOIN} \langle tname\rangle       $}
\leftline{$(8)\langle qualified\ join\rangle       ::= \langle tname\rangle         [ \text{INNER} |  \text{LEFT} | \text{RIGHT} | \text{FULL}]\ \text{JOIN}\langle tname\rangle \langle on\ clause\rangle     $}
\leftline{$(9)\langle natural\ join\rangle       ::=\langle tname\rangle        \text{NATURAL}\ \text{JOIN}\  \langle tname\rangle      $}
\leftline{$(10)\langle on\ clause\rangle       ::= \text{ON}\ \langle bvexp\rangle      $}

\leftline{$(11)\langle where\ clause\rangle       ::= \text{WHERE} \langle bvexp\rangle        $}


\leftline{$(12)\langle select\ clause\rangle       ::= \text{SELECT}\ \langle sop\rangle      |\langle af\rangle      \ \langle slist\rangle       $}
\leftline{$(13)\langle slist\rangle       ::= * | \langle cname\rangle      \ [ {,\ \langle cname\rangle       }... ]$}
\leftline{$(14)\langle sop\rangle       ::= \text{DISTINCT}\ | \text{ALL}\ $}
\leftline{$(15)\langle af\rangle       ::= \text{MAX}\ | \text{MIN}\ |\text{SUM}\ |\text{COUNT}\ |\text{AVG}\ $}


\leftline{$(16)\langle group\ by\ clause\rangle       ::= \text{GROUP}\ \text{BY}\  \langle cname\rangle      $}


\leftline{$(17)\langle having\ clause\rangle       ::= \text{HAVING} \langle bvexp\rangle        $}


\leftline{$(18)\langle order\ by\ clause\rangle       ::= \text{ORDER}\ \text{BY} \langle cname\rangle        \text{ASC} | \text{DESC}$}
\leftline{$
\begin{aligned}
(19)\langle vexp\rangle       ::= \langle nvexp\rangle\ |\ \langle svexp\rangle\       |\ \langle bvexp\rangle\ |\ \langle caseexp\rangle\ |\ \langle castexp\rangle\ |\  \langle cname\rangle\ |\ null
\end{aligned}
$}
\leftline{$(20)\langle svexp\rangle       ::= \langle concatenation\rangle\ |\ \langle character\ substring\ function\rangle\ |\ \langle trim\ function\rangle\ |\ \langle fold\rangle\ |\ \langle vexp\rangle\ $}
    \leftline{$\ \ \ \ |\ string\ literal$}
\leftline{$(21)\langle concatenation\rangle       ::= \langle svexp\rangle       || \langle svexp\rangle      $}
\leftline{$(22)\langle character\ substring\ function\rangle       ::= \text{SUBSTRING} ( \langle svexp\rangle       FROM\langle vexp\rangle       ) $}
\leftline{$(23)\langle trim\ function\rangle       ::= \text{LTRIM}|\text{RTRIM} (\langle svexp\rangle       )$}
\leftline{$(24)\langle fold\rangle       ::=  \text{UPPER}|\text{LOWER}( \langle svexp\rangle       )$}
 \leftline{$
(25)\langle nvexp\rangle       ::= \langle arithmetic\ expression\rangle\ |\ \langle modules\ expression\rangle      |\ \langle length\ expression\rangle      
$}
 \leftline{$
\ \ \ \ |\ \langle absolute\ value\ expression\rangle      |\ \langle natural\ logarithm\rangle      |\ \langle exponential\ function\rangle      |\ \langle power\ function\rangle      |\ \langle square\ root\rangle  
$}
 \leftline{$
\ \ \ \ |\ \langle floor\ function\rangle      |\ \langle ceiling\ function\rangle      |\ \langle vexp\rangle      |\ numeric\ literal
$}
\leftline{$
(26)\langle arithmetic\ expression\rangle       ::= \langle nvexp\rangle       + \langle nvexp\rangle\        |\ \langle nvexp\rangle       - \langle nvexp\rangle\ |\ \langle nvexp\rangle       * \langle nvexp\ \rangle |\ \langle nvexp\rangle       / \langle nvexp\rangle
$}
\leftline{$(27)\langle modules\ expression\rangle       ::= \text{MOD} (\langle nvexp\rangle      \ , \langle nvexp\rangle      )$}
\leftline{$(28)\langle length\ expression\rangle       ::=\{ \text{LENGTH} | \text{CHAR_LENGTH} | \text{CHARACTER_LENGTH} \}( \langle svexp\rangle      )$}
\leftline{$(29)\langle absolute\ value\ expression\rangle       ::= \text{ABS} ( \langle nvexp\rangle       )$}
\leftline{$(30)\langle natural\ logarithm\rangle       ::= \text{LN} ( \langle nvexp\rangle       )$}
\leftline{$(31)\langle exponential\ function\rangle       ::= \text{EXP} ( \langle nvexp\rangle       )$}
\leftline{$(32)\langle power\ function\rangle       ::= \text{POWER} ( \langle nvexp\rangle       , \langle nvexp\rangle       )$}
\leftline{$(33)\langle square\ root\rangle       ::= \text{SQRT} ( \langle nvexp\rangle       )$}
\leftline{$(34)\langle floor\ function\rangle       ::= \text{FLOOR} ( \langle nvexp\rangle       )$}
\leftline{$(35)\langle ceiling\ function\rangle       ::= \{ \text{CEIL} | \text{CEILING} \} ( \langle nvexp\rangle       )$}
\leftline{$
\begin{aligned}
(36)\langle bvexp\rangle       ::= \langle logical\ expression\rangle\ |\ \langle is\ expression\rangle\ |\ \langle comparison\ expression\rangle\ |\ \langle between\ expression\rangle\ 
\end{aligned}
$}
\leftline{$
\begin{aligned}
\ \ \ \ \ \langle in\ expression\rangle\ |\ \langle exists\ expression\rangle\ |\ \langle null\ expression\rangle\ |\ \langle vexp\rangle\ |\ true\ |\ false\ |\ null
\end{aligned}
$}
\leftline{$
\begin{aligned}
(37)\langle logical\ expression\rangle       ::= \langle bvexp\rangle\       \text{OR}\ \langle bvexp\rangle\ |\ \langle bvexp\rangle\ \text{AND}\ \langle bvexp\rangle\ |\ \langle bvexp\rangle\       \text{XOR}\langle bvexp\rangle\       | \text{NOT}\ \langle bvexp\rangle 
\end{aligned}
$}
\leftline{$
\begin{aligned}
(38)\langle is\ expression\rangle       ::= \langle bvexp\rangle\        \text{IS}\ [ \text{NOT} ]\ {\text{TRUE} | \text{FALSE} | \text{UNKNOWN} } \\
\end{aligned}
$}
\leftline{$
\begin{aligned}
(39)\langle comparison\ expression\rangle       ::= \langle bvexp\rangle       = | != | <  | >|> = | <= \langle bvexp\rangle\\
\end{aligned}
$}
\leftline{$
\begin{aligned}
(40)\langle between\ expression\rangle       ::= \langle nvexp\rangle        [ \text{NOT} ]\ \text{BETWEEN}\ \langle nvexp\rangle      \ \text{AND}\ \langle nvexp\rangle
\end{aligned}
$}
\leftline{$
\begin{aligned}
(41)\langle in\ expression\rangle       ::= \langle vexp\rangle      \ [ \text{NOT} ]\ \text{IN}\  \langle vlist\rangle       \\
\end{aligned}
$}
\leftline{$
\begin{aligned}
(42)\langle vlist\rangle       ::= ( \langle vexp\rangle       [ { , \langle vexp\rangle       }... ]  ) \\
\end{aligned}
$}
\leftline{$
\begin{aligned}
(43)\langle exists\ expression\rangle       ::= \text{EXISTS}\ \langle subquery\rangle      \\
\end{aligned}
$}
\leftline{$
\begin{aligned}
(44)\langle null\ expression\rangle       ::= \langle bvexp\rangle      \ \text{IS}\ [ \text{NOT} ]\ \text{NULL}\\
\end{aligned}
$}
\leftline{$(45)\langle subquery\rangle       ::= ( \langle query\ expression\rangle       )$}
\leftline{$
\begin{aligned}
(46)\langle caseexp\rangle       ::= \text{CASE}\ \text{WHEN}\ \langle vexp\rangle      \ \text{THEN}\ \langle vexp \rangle      \ \text{ELSE}\ \langle vexp\rangle
\end{aligned}
$}
\leftline{$
\begin{aligned}
(47)\langle castexp\rangle       ::= \text{CAST}(\langle vexp\rangle      \ \text{AS}\ \langle data\ type\rangle      )
\end{aligned}
$}
\leftline{$(48)\langle data\ type\rangle       ::= string\ |\ numeric\ |\ boolean      
$}
\leftline{$(50)\langle tname\rangle       ::= identifier       $}
\leftline{$(51)\langle cname\rangle       ::= identifier       $}
\end{framed}
\caption{The full list of syntax for SQL}
\label{fig:wenfa}
\end{figure*}
\begin{figure*}[tp]
\begin{framed}
\small
\leftline{\textbf{Boolean\ value\  expression($\mathcal{B}:E_B \mapsto \mathbb{B}$)} }
\leftline{$
1.\ \mathcal{B}(\langle bvexp_1\rangle) \triangleq \mathcal{B}(\langle bvexp_1\rangle       or \langle bvexp_2\rangle)|\mathcal{B}(\langle bvexp_1\rangle       and \langle bvexp_2\rangle      )|\mathcal{B}(\langle bvexp_1\rangle       xor \langle bvexp_2\rangle      )|\mathcal{B}(not \langle bvexp\rangle      )|\mathcal{B}(\langle vexp\rangle      \ is\ true)
$}
\leftline{
$
\ \ \ |\mathcal{B}(\langle bvexp\rangle      \ is\ false)|\mathcal{B}(\langle bvexp\rangle      \ is\ unknown)|\mathcal{B}(\langle bvexp\rangle      \ is\ not\ true)|\mathcal{B}(\langle bvexp\rangle      \ is\ not\ false)|\mathcal{B}(\langle bvexp\rangle      \ is\ not\ unknown)
$}
\leftline{
$
\ \ \ |\mathcal{B}(\langle nvexp_1\rangle      \ =\ \langle nvexp_2\rangle      )|\mathcal{B}(\langle nvexp_1\rangle      \ !=\ \langle nvexp_2\rangle      )|\mathcal{B}(\langle nvexp_1\rangle      \ >\ \langle nvexp_2\rangle      )|\mathcal{B}(\langle nvexp_1\rangle      \ <\ \langle nvexp_2\rangle      )
$}
\leftline{
$
\ \ \ |\mathcal{B}(\langle nvexp_1\rangle      \ >=\ \langle nvexp_2\rangle      )|\mathcal{B}(\langle nvexp_1\rangle      \ <=\ \langle nvexp_2\rangle      )|\mathcal{B}(\langle nvexp \rangle\       between\ \langle nvexp_1 \rangle\       and\  \langle nvexp_2\rangle      )|\mathcal{B}(exists\ \langle subquery\rangle      )
$}
\leftline{
$
\ \ \ |\mathcal{B}(\langle bvexp\rangle      \ is\ null)|\mathcal{B}(\langle bvexp\rangle      \ is\ not\ null)
$}
\leftline{$
2.\ \mathcal{B}(\langle bvexp_1\rangle       or \langle bvexp_2\rangle      ) \triangleq \mathcal{B}(\langle bvexp_1\rangle      ) \vee  \mathcal{B}(\langle bvexp_2\rangle      )
$}
\leftline{$
3.\ \mathcal{B}(\langle bvexp_1\rangle       and \langle bvexp_2\rangle      ) \triangleq \mathcal{B}(\langle bvexp_1\rangle      ) \wedge  \mathcal{B}(\langle bvexp_2\rangle      )
$}
\leftline{$
4.\ \mathcal{B}(\langle bvexp_1\rangle       xor \langle bvexp_2\rangle      ) \triangleq \mathcal{B}(\langle bvexp_1\rangle      ) \oplus  \mathcal{B}(\langle bvexp_2\rangle      )
$}
\leftline{$
5.\ \mathcal{B}(not \langle bvexp\rangle      ) \triangleq \neg  \mathcal{B}(\langle bvexp\rangle      )
$}
\leftline{$
\begin{aligned}
6.\ \mathcal{B}(\langle vexp\rangle      \ is\ true) \triangleq \mathcal{B}(\langle vexp\rangle      )\ =\ true
\end{aligned}
$}
\leftline{$
\begin{aligned}
7.\ \mathcal{B}(\langle bvexp\rangle      \ is\ false) \triangleq \mathcal{B}(\langle bvexp\rangle      )\ =\ false
\end{aligned}
$}
\leftline{$
\begin{aligned}
8.\ \mathcal{B}(\langle bvexp\rangle      \ is\ unknown) \triangleq \mathcal{B}(\langle bvexp\rangle      )\ =\ null
\end{aligned}
$}
\leftline{$
\begin{aligned}
9.\ \mathcal{B}(\langle bvexp\rangle      \ is\ not\ true) \triangleq \mathcal{B}(\langle bvexp\rangle      )\ \neq\ true
\end{aligned}
$}
\leftline{$
\begin{aligned}
10.\ \mathcal{B}(\langle bvexp\rangle      \ is\ not\ false) \triangleq \mathcal{B}(\langle bvexp\rangle      )\ \neq\ false
\end{aligned}
$}
\leftline{$
\begin{aligned}
11.\ \mathcal{B}(\langle bvexp\rangle      \ is\ not\ unknown) \triangleq \mathcal{B}(\langle bvexp\rangle      )\ \neq\ null
\end{aligned}
$}
\leftline{$
\begin{aligned}
12.\ \mathcal{B}(\langle nvexp_1\rangle      \ =\ \langle nvexp_2\rangle      ) \triangleq \mathcal{N}(\langle nvexp_1\rangle      ) =  \mathcal{N}(\langle nvexp_2\rangle      )
\end{aligned}
$}
\leftline{$
\begin{aligned}
13.\ \mathcal{B}(\langle nvexp_1\rangle      \ !=\ \langle nvexp_2\rangle      ) \triangleq \mathcal{N}(\langle nvexp_1\rangle      ) \neq  \mathcal{N}(\langle nvexp_2\rangle      ) 
\end{aligned}
$}
\leftline{$
\begin{aligned}
14.\ \mathcal{B}(\langle nvexp_1\rangle      \ >      \ \langle nvexp_2\rangle      ) \triangleq \mathcal{N}(\langle nvexp_1\rangle      ) >        \mathcal{N}(\langle nvexp_2\rangle      )
\end{aligned}
$}
\leftline{$
\begin{aligned}
15.\ \mathcal{B}(\langle nvexp_1\rangle      \ < \ \langle nvexp_2\rangle      ) \triangleq \mathcal{N}(\langle nvexp_1\rangle      ) <   \mathcal{N}(\langle nvexp_2\rangle      )
\end{aligned}
$}
\leftline{$
\begin{aligned}
16.\ \mathcal{B}(\langle nvexp_1\rangle      \ >=\ \langle nvexp_2\rangle      ) \triangleq \mathcal{N}(\langle nvexp_1\rangle      ) \geq  \mathcal{N}(\langle nvexp_2\rangle      )
\end{aligned}
$}
\leftline{$
\begin{aligned}
17.\ \mathcal{B}(\langle nvexp_1\rangle      \ <=\ \langle nvexp_2\rangle      ) \triangleq \mathcal{N}(\langle nvexp_1\rangle      ) \leq  \mathcal{N}(\langle nvexp_2\rangle      )
\end{aligned}
$}
\leftline{$
\begin{aligned}
18.\ \mathcal{B}(\langle nvexp \rangle\       between\ \langle nvexp_1 \rangle\       and\  \langle nvexp_2\rangle      ) \triangleq (\mathcal{N}(\langle nvexp\rangle      )\leq \mathcal{N}(\langle nvexp_2\rangle      )) \wedge(\mathcal{N}(\langle nvexp\rangle      )\geq\mathcal{N}(\langle nvexp_1\rangle      ))\\
\end{aligned}
$}
\leftline{$
\begin{aligned}
19.\ \mathcal{B}(\langle vexp\rangle\        in\ ( \langle vexp_1\rangle [ { , \langle vexp_2\rangle }... ]  )      ) \triangleq \langle vexp\rangle  \in  ( \langle vexp_1\rangle [ { , \langle vexp_2\rangle }... ]  )\\
\end{aligned}
$}
\leftline{$
\begin{aligned}
20.\ \mathcal{B}(exists\ \langle subquery\rangle      ) \triangleq \mathcal{H}[\![\langle subquery\rangle      ]\!] \neq \oslash\\
\end{aligned}
$}
\leftline{$
\begin{aligned}
21.\ \mathcal{B}(\langle bvexp\rangle      \ is\ null) \triangleq \mathcal{B}(\langle bvexp\rangle      )\ =\ null
\end{aligned}
$}
\leftline{$
\begin{aligned}
22.\ \mathcal{B}(\langle bvexp\rangle      \ is\ not\ null) \triangleq \mathcal{B}(\langle bvexp\rangle      )\ \neq\ null
\end{aligned}
$}
\leftline{$
\begin{aligned}
23.\ \mathcal{B}(\langle vexp\rangle      ) \triangleq \mathcal{B}(\langle nvexp\rangle      )|\mathcal{B}(\langle svexp\rangle      )|\mathcal{B}(\langle caseexp\rangle      )|\mathcal{B}(\langle castexp\rangle      )|\mathcal{B}(\langle cname\rangle      )|\mathcal{B}(null)
\end{aligned}
$}
\leftline{$
\begin{aligned}
24.\ \mathcal{B}(\langle nvexp\rangle      ) \triangleq \mathcal{B}(cast(\langle nvexp\rangle      \ as\ boolean))
\end{aligned}
$}
\leftline{$
\begin{aligned}
25.\ \mathcal{B}(\langle svexp\rangle      ) \triangleq \mathcal{B}(cast(\langle svexp\rangle      \ as\ boolean))
\end{aligned}
$}
\leftline{$
\begin{aligned}
26.\ \mathcal{B}(caseexp      )\triangleq \mathcal{B}(case\ when\ \langle bvexp\rangle      \ then\ \langle vexp_1\rangle      \ else\ \langle vexp_2\rangle      )
\end{aligned}
$}
\leftline{$
27.\ \mathcal{B}(case\ when\ \langle bvexp\rangle      \ then\ \langle vexp_1\rangle      \ else\ \langle vexp_2\rangle      )\triangleq \left\{
\begin{aligned}
&\mathcal{B}(\langle vexp_1\rangle      ); \ \ \ \ \ \mathcal{B}(\langle bvexp\rangle)=true\\
&\mathcal{B}(\langle vexp_2\rangle      );\ \ \ \ \  \mathcal{B}(\langle bvexp\rangle)=false
\end{aligned}
\right.$}
\leftline{$
\begin{aligned}
28.\ \mathcal{B}(\langle castexp\rangle      ) \triangleq \mathcal{B}(cast(\langle nvexp\rangle\       as\ boolean      ))|\mathcal{B}(cast(\langle svexp\rangle\       as\ boolean      ))
\end{aligned}
$}
\leftline{$
29.\ \mathcal{B}(cast(\langle nvexp\rangle\       as\ boolean      ))\triangleq \left\{
\begin{aligned}
&null;\ \ \ \ \  \mathcal{N}(\langle nvexp\rangle      )=null\\
&false;\ \ \ \ \  \mathcal{N}(\langle nvexp\rangle      )=0\\
&true;\ \ \ \ \  otherwise\\
\end{aligned}
\right.$}
\leftline{$
30.\ \mathcal{B}(cast(\langle svexp\rangle\       as\ boolean      ))\triangleq \left\{
\begin{aligned}
&null;\ \ \ \ \  \mathcal{S}(\langle svexp\rangle      )=null\\
&false;\ \ \ \ \  \mathcal{S}(\langle svexp\rangle      )=``0"|``false"|``"\\
&true;\ \ \ \ \  otherwise\\
\end{aligned}
\right.$}
\leftline{$
\begin{aligned}
31.\ \mathcal{B}(\langle cname\rangle      ) \triangleq  \mathcal{B}(cast(S(\langle cname\rangle      )\ as\ boolean))
\end{aligned}
$}
\leftline{$
32.\ \mathcal{B}(true) \triangleq  true
$}
\leftline{$
33.\ \mathcal{B}(false) \triangleq  false
$}
\leftline{$
34.\ \mathcal{B}(null) \triangleq  null
$}
~\\
\end{framed}
\caption{The full list semantic definition of boolean expression }
\label{fig:boolfullsemantic}
\end{figure*}

\begin{figure*}[tp]
\begin{framed}
\small
\leftline{\textbf{Numeric\ value\  expression($\mathcal{N}:E_N \mapsto \mathbb{N}$)}}
\leftline{$
\begin{aligned}
35.\ \mathcal{N}(\langle nvexp\rangle) \triangleq \mathcal{N}(\langle nvexp_1\rangle       + \langle nvexp_2\rangle      )|\mathcal{N}(\langle nvexp_1\rangle       - \langle nvexp_2\rangle      )|\mathcal{N}(\langle nvexp_1\rangle       * \langle nvexp_2\rangle      )|\mathcal{N}(\langle nvexp_1\rangle       / \langle nvexp_2\rangle      )
\end{aligned}
$}
\leftline{$
\begin{aligned}
\ \ \ \ |\mathcal{N}(length | char\_length | charactor\_length(\langle svexp\rangle      )|\mathcal{N}(mod(\langle nvexp_1\rangle       , \langle nvexp_2\rangle      )|\mathcal{N}(abs(\langle nvexp\rangle      )|\mathcal{N}(ln(\langle nvexp\rangle      )
\end{aligned}
$}
\leftline{$
\begin{aligned}
\ \ \ \ |\mathcal{N}(exp(\langle nvexp\rangle      ) |\mathcal{N}(power(\langle nvexp_1\rangle      ,\langle nvexp_2\rangle      )|\mathcal{N}(sqrt(\langle nvexp\rangle      )|\mathcal{N}(floor(\langle nvexp\rangle      )|\mathcal{N}(ceil|ceiling(\langle nvexp\rangle      )
\end{aligned}
$}
\leftline{$
\begin{aligned}
36.\ \mathcal{N}(\langle nvexp_1\rangle       + \langle nvexp_2\rangle      ) \triangleq \mathcal{N}(\langle nvexp_1\rangle      ) +  \mathcal{N}(\langle nvexp_2\rangle      )
\end{aligned}
$}
\leftline{$
\begin{aligned}
37.\ \mathcal{N}(\langle nvexp_1\rangle       - \langle nvexp_2\rangle      ) \triangleq \mathcal{N}(\langle nvexp_1\rangle      ) -  \mathcal{N}(\langle nvexp_2\rangle      )
\end{aligned}
$}
\leftline{$
\begin{aligned}
38.\ \mathcal{N}(\langle nvexp_1\rangle       * \langle nvexp_2\rangle      ) \triangleq \mathcal{N}(\langle nvexp_1\rangle      ) *  \mathcal{N}(\langle nvexp_2\rangle      )
\end{aligned}
$}
\leftline{$
\begin{aligned}
39.\ \mathcal{N}(\langle vexp_1\rangle       / \langle nvexp_2\rangle      ) \triangleq \mathcal{N}(\langle nvexp_1\rangle      ) /  \mathcal{N}(\langle nvexp_2\rangle      )
\end{aligned}
$}
\leftline{$
\begin{aligned}
40.\ \mathcal{N}(length | char\_length | charactor\_length(\langle svexp\rangle      ) \triangleq len(\mathcal{S}(\langle svexp\rangle      ))
\end{aligned}
$}
\leftline{$
\begin{aligned}
41.\ \mathcal{N}(mod(\langle nvexp_1\rangle       , \langle nvexp_2\rangle      ) \triangleq   (\mathcal{N}(\langle nvexp_1\rangle      )\%\mathcal{N}(\langle nvexp_2\rangle      )
\end{aligned}
$}
\leftline{$
\begin{aligned}
42.\ \mathcal{N}(abs(\langle nvexp\rangle      ) \triangleq |\mathcal{N}(\langle nvexp\rangle)|
\end{aligned}
$}
\leftline{$
\begin{aligned}
43.\ \mathcal{N}(ln(\langle nvexp\rangle      ) \triangleq ln(\mathcal{N}(\langle nvexp\rangle      ))\\
\end{aligned}
$}
\leftline{$
\begin{aligned}
44.\ \mathcal{N}(exp(\langle nvexp\rangle      ) \triangleq  e^{\mathcal{N}(\langle nvexp\rangle      )}\\
\end{aligned}
$}
\leftline{$
\begin{aligned}
45.\ \mathcal{N}(power(\langle nvexp_1\rangle      ,\langle nvexp_2\rangle      ) \triangleq  \mathcal{N}(\langle nvexp_1\rangle      )^{\mathcal{N}(\langle nvexp_2\rangle      )} 
\end{aligned}
$}
\leftline{$
\begin{aligned}
46.\ \mathcal{N}(sqrt(\langle nvexp\rangle      ) \triangleq \sqrt{\mathcal{N}(\langle nvexp\rangle      )} 
\end{aligned}
$}
\leftline{$
\begin{aligned}
47.\ \mathcal{N}(floor(\langle nvexp\rangle      ) \triangleq \lfloor \mathcal{N}(\langle nvexp\rangle      ) \rfloor
\end{aligned}
$}
\leftline{$
\begin{aligned}
48.\ \mathcal{N}(ceil|ceiling(\langle nvexp\rangle      ) \triangleq \lceil \mathcal{N}(\langle nvexp\rangle      ) \rceil
\end{aligned}
$}
\leftline{$
\begin{aligned}
49.\ \mathcal{N}(\langle vexp\rangle      ) \triangleq \mathcal{N}(\langle bvexp\rangle      )|\mathcal{N}(\langle svexp\rangle      )|\mathcal{N}(\langle caseexp\rangle      )|\mathcal{N}(\langle castexp\rangle      )|\mathcal{N}(\langle cname\rangle      )|\mathcal{N}(null      )
\end{aligned}
$}
\leftline{$
\begin{aligned}
50.\ \mathcal{N}(\langle bvexp\rangle      ) \triangleq \mathcal{N}(cast(\langle bvexp\rangle      \ as\ numeric))
\end{aligned}
$}
\leftline{$
\begin{aligned}
51.\ \mathcal{N}(\langle svexp\rangle      ) \triangleq \mathcal{N}(cast(\langle svexp\rangle      \ as\ numeric))
\end{aligned}
$}
\leftline{$
\begin{aligned}
52.\ \mathcal{N}(\langle cname\rangle      ) \triangleq \mathcal{N}(cast(S(\langle cname\rangle      )\ as\ numeric))
\end{aligned}
$}
\leftline{$
\begin{aligned}
53.\ \mathcal{N}(caseexp      )\triangleq \mathcal{N}(case\ when\ \langle bvexp\rangle      \ then\ \langle vexp_1\rangle      \ else\ \langle vexp_2\rangle      )
\end{aligned}
$}
\leftline{$
54.\ \mathcal{N}(case\ when\ \langle bvexp\rangle      \ then\ \langle vexp_1\rangle      \ else\ \langle vexp_2\rangle      )\triangleq \left\{
\begin{aligned}
&\mathcal{N}(\langle vexp_1\rangle      ); \ \ \ \ \ \mathcal{B}(\langle bvexp\rangle)=true\\
&\mathcal{N}(\langle vexp_2\rangle      );\ \ \ \ \ \mathcal{B}(\langle bvexp\rangle)=false
\end{aligned}
\right.$}
\leftline{$
\begin{aligned}
55.\ \mathcal{N}(\langle castexp\rangle      ) \triangleq \mathcal{N}(cast(\langle bvexp\rangle\       as\ numeric      ))|\mathcal{N}(cast(\langle svexp\rangle\       as\ numeric      ))
\end{aligned}
$}
\leftline{$
56.\ \mathcal{N}(cast(\langle bvexp\rangle\       as\ numeric      ))\triangleq \left\{
\begin{aligned}
&1; \ \ \ \ \ \mathcal{B}(\langle bvexp\rangle      )=true\\
&0;\ \ \ \ \  \mathcal{B}(\langle bvexp\rangle      )=false\\
&null;\ \ \ \ \  \mathcal{B}(\langle bvexp\rangle      )=null\\
\end{aligned}
\right.$}
\leftline{$
57.\ \mathcal{N}(cast(\langle svexp\rangle\       as\ numeric      ))\triangleq \left\{
\begin{aligned}
&null;\ \ \ \ \  \mathcal{S}(\langle svexp\rangle      )=null\\
&str2num(\mathcal{S}(\langle svexp\rangle      )); \ \ \ \ \ otherwise\\
\end{aligned}
\right.$}
\leftline{$
58.\ \mathcal{N}(numeric\ literal)  \triangleq numeric\ literal
$}
\leftline{$
59.\ \mathcal{N}(null) \triangleq  null
$}
\leftline{\textbf{String\ value\  expression($\mathcal{S}:E_S \mapsto \mathbb{S}$)}}
\leftline{$60.\ \mathcal{S}(\langle svexp\rangle) \triangleq \mathcal{S}(\langle svexp_1\rangle       || \langle svexp_2\rangle      )|\mathcal{S}(substring(\langle svexp_1\rangle      \ in\ \langle nvexp_2\rangle      ))|\mathcal{S}(ltrim(\langle svexp\rangle      ))|\mathcal{S}(rtrim(\langle svexp\rangle      ))
$}
\leftline{$\ \ \ \ |\mathcal{S}(lower(\langle svexp\rangle      ))|\mathcal{S}(upper(\langle svexp\rangle      ))
$}
\leftline{$61.\ \mathcal{S}(\langle svexp_1\rangle       || \langle svexp_2\rangle      ) \triangleq \mathcal{S}(\langle svexp_1\rangle      )|| \mathcal{S}(\langle svexp_2\rangle      )
$}
\leftline{$62.\ \mathcal{S}(substring(\langle svexp_1\rangle      \ in\ \langle nvexp_2\rangle      )) \triangleq  \mathcal{S}(\langle svexp_1\rangle      )[\mathcal{N}(\langle nvexp_2\rangle      ),len(\mathcal{S}(\langle svexp_1\rangle      ))]
$}
\leftline{$63.\ \mathcal{S}(lower(\langle svexp\rangle      )) \triangleq s_{lower}  
$}
\leftline{$\ \ \ \ s_{lower}:  (len(s_{lower})=len(\mathcal{S}(\langle svexp\rangle      )))\wedge (\forall i \in (0,len(\mathcal{S}(\langle svexp\rangle))),s_{lower}[i]=\mathcal{S}(\langle svexp\rangle      )[i]+32     
$}
\leftline{$64.\ \mathcal{S}(upper(\langle svexp\rangle      )) \triangleq s_{upper}   
$}
\leftline{$\ \ \ \ s_{upper}:(len(s_{upper})=len(\mathcal{S}(\langle svexp\rangle      )))\wedge (\forall i \in (0,len(\mathcal{S}(\langle svexp\rangle))),s_{upper}[i]=\mathcal{S}(\langle svexp\rangle      )[i]-32      
$}
\leftline{$
\begin{aligned}
65.\ \mathcal{S}(\langle vexp\rangle      ) \triangleq \mathcal{S}(\langle bvexp\rangle      )|\mathcal{S}(\langle nvexp\rangle      )|\mathcal{S}(\langle caseexp\rangle      )|\mathcal{S}(\langle castexp\rangle      )|\mathcal{S}(\langle cname\rangle      )|\mathcal{S}(null      )
\end{aligned}
$}
\leftline{$
\begin{aligned}
66.\ \mathcal{S}(\langle bvexp\rangle      ) \triangleq \mathcal{S}(cast(\langle bvexp\rangle      \ as\ string))\ 
\end{aligned}
$}
\leftline{$
\begin{aligned}
67.\ \mathcal{S}(\langle nvexp\rangle      ) \triangleq \mathcal{S}(cast(\langle nvexp\rangle      )\ as\ string)\ 
\end{aligned}
$}
\leftline{$
\begin{aligned}
68.\ \mathcal{S}(caseexp      )\triangleq \mathcal{S}(case\ when\ \langle bvexp\rangle      \ then\ \langle vexp_1\rangle      \ else\ \langle vexp_2\rangle      )
\end{aligned}
$}
\leftline{$
69.\ \mathcal{S}(case\ when\ \langle bvexp\rangle      \ then\ \langle vexp_1\rangle      \ else\ \langle vexp_2\rangle      )\triangleq \left\{
\begin{aligned}
&\mathcal{S}(\langle vexp_1\rangle      ); \ \ \ \ \ \mathcal{B}(\langle bvexp\rangle)=true\\
&\mathcal{S}(\langle vexp_2\rangle      );\ \ \ \ \  \mathcal{B}(\langle bvexp\rangle)=false
\end{aligned}
\right.$}
\leftline{$
\begin{aligned}
70.\ \mathcal{S}(\langle castexp\rangle      ) \triangleq \mathcal{S}(cast(\langle bvexp\rangle\       as\ string      ))|\mathcal{S}(cast(\langle nvexp\rangle\       as\ string      ))
\end{aligned}
$}
\leftline{$
71.\ \mathcal{S}(cast(\langle bvexp\rangle\       as\ string      ))\triangleq \left\{
\begin{aligned}
&1; \ \ \ \ \ \mathcal{B}(\langle bvexp\rangle      )=true\\
&0;\ \ \ \ \  \mathcal{B}(\langle bvexp\rangle      )=false\\
&null;\ \ \ \ \  \mathcal{B}(\langle bvexp\rangle      )=null\\
\end{aligned}
\right.$}
\leftline{$
72.\ \mathcal{S}(cast(\langle nvexp\rangle\       as\ string      ))\triangleq \left\{
\begin{aligned}
&null;\ \ \ \ \  \mathcal{N}(\langle nvexp\rangle      )=null\\
&num2str(\mathcal{N}(\langle nvexp\rangle      )); \ \ \ \ \ otherwise\\
\end{aligned}
\right.$}
\leftline{$
\begin{aligned}
73.\ \mathcal{S}(\langle cname\rangle      ) \triangleq \mathcal{S}(cast(S(\langle cname\rangle      )\ as\ string))\ 
\end{aligned}
$}
\leftline{$74.\ \mathcal{S}(string\ literal) \triangleq string\ literal      
$}
\leftline{$
75.\ \mathcal{S}(null) \triangleq  null
$}
\end{framed}
\caption{The full list of semantic definition for numeric expression and string expression}
\label{fig:numstrfullsemantic}
\end{figure*}
\begin{figure*}[tp]
\begin{framed}
\small
\leftline{\textbf{Keyword operation($\mathcal{C}:\{L,OP\} \mapsto T$)}}
\leftline{\textbf{Join operation}}
\leftline{$
\begin{aligned}
76.\ \mathcal{C}[\![\{[T_1,T_2],\ natural\ join\}]\!] \triangleq \{\alpha_1 \circ \alpha_2| \alpha_1 \in T_1,\alpha_2 \in T_2, \overline{\beta}_I = T_1.\overline{\beta}\cap T_2.\overline{\beta}\}
\end{aligned}
$}
\leftline{$
\ \ \ \ \alpha_1 \circ \alpha_2 \triangleq \left\{
\begin{aligned}
&\alpha_1\ \bowtie \alpha_2; \ \ \ \ \ (\overline{\beta}_I \neq \emptyset) \wedge(\pi_{\overline{\beta}_I}(\{\alpha_1\}) = \pi_{\overline{\beta}_I}(\{\alpha_2\}))\\
&skip;\ \ \ \ \ \ \ \ \ \ \ \ \ \ \ \ \ \ \ \ \ \ \ \ \ \ \ \ \ \ \ \ \ \  otherwise
\end{aligned}
\right.$}
\leftline{$
\begin{aligned}
77.\ \mathcal{C}[\![\{[T_1,T_2],\ left\ join\}]\!] \triangleq \{ \alpha_1 \bullet \alpha_2| \alpha_1 \in T_1,\alpha_2 \in T_2, \overline{\beta}_I = T_1.\overline{\beta}\cap T_2.\overline{\beta}\} \\  
\end{aligned}
$}
\leftline{$
\ \ \ \ \alpha_1 \bullet \alpha_2 \triangleq \left\{
\begin{aligned}
&\alpha_1\ \bowtie \alpha_2; \ \ \ \ \ \ \ \ \ \ \ \ \ \ \ (\overline{\beta}_I \neq \emptyset)\wedge(\pi_{\overline{\beta}_I}(\{\alpha_1\}) = \pi_{\overline{\beta}_I}(\{\alpha_2\}))\\
&\alpha_1 \bowtie [null,...,null]_{|\{\alpha_2\}.\overline{\beta}|-|\overline{\beta}_I|};\ \ \ \ (\overline{\beta}_I \neq \emptyset)\wedge(\pi_{\overline{\beta}_I}(\{\alpha_1\}) \neq \pi_{\overline{\beta}_I}(\{\alpha_2\}))\\
&skip;\ \ \ \ otherwise
\end{aligned}
\right.$}
\leftline{$
\begin{aligned}
78.\ \mathcal{C}[\![\{[T_1,T_2],\ right\ join\}]\!] \triangleq \{ \alpha_1 \bullet \alpha_2| \alpha_1 \in T_1,\alpha_2 \in T_2, \overline{\beta}_I = T_1.\overline{\beta}\cap T_2.\overline{\beta}\} \\  
\end{aligned}
$}
\leftline{$
\ \ \ \ \alpha_1 \bullet \alpha_2 \triangleq \left\{
\begin{aligned}
&\alpha_1\ \bowtie \alpha_2; \ \ \ \ \ \ \ \ \ \ \ \ \ \ \ (\overline{\beta}_I \neq \emptyset)\wedge(\pi_{\overline{\beta}_I}(\{\alpha_1\}) = \pi_{\overline{\beta}_I}(\{\alpha_2\}))\\
& [null,...,null]_{|\{\alpha_1\}.\overline{\beta}|-|\overline{\beta}_I|} \bowtie \alpha_2;\ \ \ \ (\overline{\beta}_I \neq \emptyset)\wedge(\pi_{\overline{\beta}_I}(\{\alpha_1\}) \neq \pi_{\overline{\beta}_I}(\{\alpha_2\}))\\
&skip;\ \ \ \ otherwise
\end{aligned}
\right.$}
\leftline{$
\begin{aligned}
79.\ \mathcal{C}[\![\{[T_1,T_2],\ cross\ join\}]\!] \triangleq  \{\alpha_1 \times \alpha_2| \alpha_1 \in T_1,\alpha_2 \in T_2\}
\end{aligned}
$}
\leftline{$
\begin{aligned}
80.\ \mathcal{C}[\![\{[T_1,T_2],\ inner\ join\}]\!] \triangleq  \mathcal{C}[\![\{[T_1,T_2],\ cross\ join\}]\!]
\end{aligned}
$}
~\\
\leftline{\textbf{Collection operation}}
\leftline{$
81.\ \mathcal{C}[\![\{[T_1,T_2],\ union\}]\!] \triangleq \{\alpha_{u(T_1,T_2)}| \xi_{\alpha_{u(T_1,T_2)}}(T)=1\} 
$}
\leftline{$
\ \ \ \ \mathcal{C}[\![\{[T_1,T_2],\ union\ all\}]\!] \triangleq \{\alpha_{u(T_1,T_2)}\} 
$}
\leftline{$
\ \ \ \ \alpha_{u(T_1,T_2)} \triangleq \left\{
\begin{aligned}
&\alpha; \ \ \ \ \ \ \ \ \ \ \ (\alpha \in T_1)\vee(\alpha \in T_2)\\
&skip;\ \ \ \ \ \ \ \  otherwise
\end{aligned}
\right.$}
~\\
\leftline{$
82.\ \mathcal{C}[\![\{[T_1,T_2],\ intersect\}]\!] = \{\alpha_{i(T_1,T_2)} | \xi_{\alpha_{i(T_1,T_2)}}(T)=1\}
$}
\leftline{$
\ \ \ \ \mathcal{C}[\![\{[T_1,T_2],\ intersect\ all\}]\!] = \{\alpha_{i(T_1,T_2)} | \xi_{\alpha_{i(T_1,T_2)}}(T)=min(\xi_{\alpha_{i(T_1,T_2)}}(T_1),\xi_{\alpha_{i(T_1,T_2)}}(T_2))\}
$}
\leftline{$
\ \ \ \ \alpha_{i(T_1,T_2))}=\left\{
\begin{aligned}
&\alpha; \ \ \ \ \ \ \ \ \ \ \ (\alpha \in T_1)\wedge(\alpha \in T_2)\\
&skip;\ \ \ \ \ \ \ \  otherwise
\end{aligned}
\right.$}
~\\
\leftline{$
83.\ \mathcal{C}[\![\{[T_1,T_2],\ except\}]\!] = \{\alpha_{e(T_1,T_2} | \xi_{\alpha_{e(T_1,T_2}}(T)=1\}
$}
\leftline{$
\ \ \ \ \alpha_{e(T_1,T_2))}=\left\{
\begin{aligned}
&\alpha; \ \ \ \ \ \ \ \ \ \ \ (\alpha \in T_1)\wedge(\alpha \notin T_2)\\
&skip;\ \ \ \ \ \ \ \  otherwise
\end{aligned}
\right.$}
\leftline{$
\ \ \ \ \mathcal{C}[\![\{[T_1,T_2],\ except\ all\}]\!] = \{\alpha_{ea(T_1,T_2)} | \xi_{\alpha_{ea(T_1,T_2)}}(T)=max(0,\xi_{\alpha_{ea(T_1,T_2)}}(T_1)-\xi_{\alpha_{ea(T_1,T_2)}}(T_2))\}
$}
\leftline{$
\ \ \ \ \alpha_{ea(T_1,T_2)}=\left\{
\begin{aligned}
&\alpha; \ \ \ \ \ \ \ \ \ \ \ \xi_{\alpha}(T_1)>\xi_{\alpha}(T_2)\\
&skip;\ \ \ \ \ \ \ \  otherwise
\end{aligned}
\right.$}
~\\
\leftline{\textbf{Filter operation}}
\leftline{$
\begin{aligned}
84.\ \mathcal{C}[\![\{[T],\ distinct\}]\!] \triangleq \{\alpha| 
(\forall \alpha \in T,\ \xi_{\alpha} (T_1)=1) \wedge(\forall \alpha \in T_1,\ \alpha \in T )
\}
\end{aligned}
$}
\leftline{$
\begin{aligned}
85.\ \mathcal{C}[\![\{[T],\ all\}]\!] \triangleq T
\end{aligned}
$}
~\\
\leftline{\textbf{Aggregation operation}}
\leftline{$
\begin{aligned}
86.\ \mathcal{C}[\![\{[T],\ max\}]\!] \triangleq \{v| 
(v \in \pi_{T.\overline{\beta}}(T)) \wedge(\forall v_1 \in \pi_{T.\overline{\beta}}(T), \sigma_{(v_1 >       v)}(\pi_{T.\overline{\beta}}(T))=\oslash)
\}
\end{aligned}
$}
\leftline{$
\begin{aligned}
87.\ \mathcal{C}[\![\{[T],\ min\}]\!] \triangleq \{v| 
(v \in \pi_{T.\overline{\beta}}(T)) \wedge(\forall v_1 \in \pi_{T.\overline{\beta}}(T), \sigma_{(v_1 <  v)}(\pi_{T.\overline{\beta}}(T))=\oslash)
\}
\end{aligned}
$}
\leftline{$
\begin{aligned}
88.\ \mathcal{C}[\![\{[T],\ sum\}]\!] \triangleq \{v|
v= \sum_{v_1 \in \pi_{T.\overline{\beta}}(T)}v_1
\}
\end{aligned}
$}
\leftline{$
\begin{aligned}
89.\ \mathcal{C}[\![\{[T],\ count\}]\!] \triangleq \{v|
v=\sum_{\alpha \in \pi_{T.\overline{\beta}}(T)}(\xi_{\alpha} (T))
\}
\end{aligned}
$}
\leftline{$
\begin{aligned}
90.\ \mathcal{C}[\![\{[T],\ avg\}]\!] \triangleq \{v|
v= \sum_{v_1 \in \pi_{T.\overline{\beta}}(T)}v_1/\sum_{\alpha \in \pi_{T.\overline{\beta}}(T)}(\xi_{\alpha} (T))
\}
\end{aligned}
$}
\end{framed}
\caption{The full list of semantic definitions for join operation, collection operation, filtering operation, and aggregation operation}
\label{fig:joinfullsemantic}
\end{figure*}
\begin{figure*}[tp]
\begin{framed}
\small
\leftline{\textbf{Keyword operation($\mathcal{H}:S \mapsto (A,T)$)}}
\leftline{\textbf{Single keyword operation}}
\leftline{$
\begin{aligned}
91.\ \mathcal{H}[\![from \langle tname\rangle      ]\!] \triangleq 
(A, A(\langle tname\rangle      ))
\end{aligned}
$}
\leftline{$
\begin{aligned}
92.\ \mathcal{H}[\![from \langle tname_1\rangle      , \langle tname_2\rangle      ]\!] \triangleq (A, \mathcal{C}[\![{[A(\langle tname_1\rangle      ),A(\langle tname_2\rangle      )],\ cross\ join\ }]\!] )
\end{aligned}
$}
\leftline{$
\begin{aligned}
93.\ \mathcal{H}[\![select\ *]\!] \triangleq (\{T\}, \pi_{T.\overline{\beta}}(T))
\end{aligned}
$}
\leftline{$
\begin{aligned}
94.\ \mathcal{H}[\![select\ \langle cname\rangle      \ [ {,\ \langle cname\rangle       }... ]]\!] \triangleq (\{T\}, \pi_{\langle cname\rangle      \ [ {,\ \langle cname\rangle       }... ]}(T))
\end{aligned}
$}
\leftline{$
\begin{aligned}
95.\ \mathcal{H}[\![on\ \langle bvexp\rangle      ]\!] \triangleq (\{T\}, \sigma_{\mathcal{B}(\langle bvexp\rangle      )}(T))
\end{aligned}
$}
\leftline{$
\begin{aligned}
96.\ \mathcal{H}[\![where\ \langle bvexp\rangle      ]\!] \triangleq (\{T\}, \sigma_{\mathcal{B}(\langle bvexp\rangle      )}(T))
\end{aligned}
$}
\leftline{$
\begin{aligned}
97.\ \mathcal{H}[\![group\ by\ \langle cname\rangle      ]\!] \triangleq (\{T\}, (\widehat{\alpha_1},...,\widehat{\alpha_k}):\forall \widehat{\alpha_p}\in\ (\widehat{\alpha_1},...,\widehat{\alpha_k}), 
 \forall\ v_{ij}\in\ \pi_{\langle cname\rangle      }(\widehat{\alpha_p}),(v_{ij}=v_{i1}) )
\end{aligned}
$}
\leftline{$
\begin{aligned}
98.\ \mathcal{H}[\![having\ \langle bvexp\rangle      ]\!] \triangleq (\{T\},\sigma_{\mathcal{B}(\langle bvexp\rangle      )}(T))
\end{aligned}
$}
\leftline{$
\begin{aligned}
99.\ \mathcal{H}[\![order\ by\ \langle cname\rangle\       asc]\!] \triangleq (\{T\}, T_1)\ where\  (\forall \alpha \in T_1,\ \xi_{\alpha} (T_1)=\xi_{\alpha} (T)) 
\wedge(\forall \alpha \in T,
\end{aligned}
$}
\leftline{
$
\begin{aligned}
     \ \ \ \ \xi_{\alpha} (T)=\xi_{\alpha} (T_1))\wedge(\forall v_i, v_j\in \sigma_{T_1.{\langle cname\rangle      }}(T_1), i> j \text{\space} \text{iff} \ v_i >= v_{j} )
\end{aligned}
$}
\leftline{$
\begin{aligned}
100.\ \mathcal{H}[\![order\ by\ \langle cname\rangle\       desc]\!] \triangleq (\{T\}, T_1)\ where\  (\forall \alpha \in T_1,\ \xi_{\alpha} (T_1)=\xi_{\alpha} (T)) 
\wedge(\forall \alpha 
\end{aligned}
$}
\leftline{
$
\begin{aligned}
    \ \ \ \ \in T,\xi_{\alpha} (T)=\xi_{\alpha} (T_1))\wedge(\forall v_i, v_j\in \sigma_{T_1.{\langle cname\rangle      }}(T_1), i< j \text{\space} \text{iff} \ v_i <= v_{j} )
\end{aligned}
$}
\leftline{$
\begin{aligned}
101.\ \mathcal{H}[\![\langle subquery\rangle      ]\!] \triangleq\mathcal{H}[\![\langle query\ expression\rangle      ]\!] 
\end{aligned}
$}
~\\
\leftline{\textbf{Composite keyword operation}}
\leftline{$
\begin{aligned}
102.\ \diamond:\ (1)
\frac{\mathcal{H}[\![expression_1]\!] \triangleq (A,T),\ \mathcal{H}[\![expression_2]\!] \triangleq (\{T\},T')}{\mathcal{H}[\![expression_1]\!] \diamond \mathcal{H}[\![expression_2]\!] \triangleq (A,T')}
\end{aligned}
$}
\leftline{$
\begin{aligned}
\ \ \ \ \ \ \ \ \ \ \ \ (2)
\frac{\mathcal{H}[\![expression]\!] \triangleq (A,T),\ \mathcal{C}[\![\{\{T\},OP\}]\!] \triangleq T'}{\mathcal{H}[\![expression]\!] \diamond \mathcal{C}[\![\{\{T\},OP\}]\!] \triangleq (A,T')}
\end{aligned}
$}
\leftline{$
\begin{aligned}
\ \ \ \ \ \ \ \ \ \ \ \ (3)
\frac{\mathcal{C}[\![\{L,OP\}]\!] \triangleq T,\ \mathcal{H}[\![expression]\!] \triangleq (\{T\},T')}{\mathcal{C}[\![\{L,OP\}]\!] \diamond \mathcal{H}[\![expression]\!]  \triangleq (L,T')}
\end{aligned}
$}
\leftline{$
\begin{aligned}
\ \ \ \ \ \ \ \ \ \ \ \ (4)
\frac{\mathcal{H}[\![expression_1]\!] \triangleq (A,T_1),\ \mathcal{H}[\![expression_2]\!] \triangleq (A,T_2),\ \mathcal{C}[\![\{\{T_1,T_2\},OP\}]\!] \triangleq T_3}{(\mathcal{H}[\![expression_1]\!],\mathcal{H}[\![expression_2]\!]) \diamond \mathcal{C}[\![\{\{T_1,T_2\},OP\}]\!] \triangleq (A,T_3)}
\end{aligned}
$}
\leftline{$
\begin{aligned}
103.\ \mathcal{H}[\![\langle queryexp\rangle]\!]=
\end{aligned}
$}
\leftline{$
\begin{aligned}
\ \ \ \ \ \ \mathcal{H}[\![select\ \textbf{[}\langle sop\rangle|\langle af\rangle\textbf{]}\ \langle cname_1\rangle     \textbf{[}, \langle cname_2\rangle ...\textbf{]}    \ from \langle tname_1\rangle       \textbf{[} { , \langle tname_2\rangle       }... \textbf{]}|from\ \langle tname_1 \rangle\       natural/cross\ join\ \langle tname_2\rangle
\end{aligned}
$}
\leftline{$
\begin{aligned}
\ \ \ \ \ \ |from\ \langle tname_1 \rangle\       left/right/full/inner\ join\ \langle tname_2\rangle\ on\langle bvexp\rangle\ \textbf{[}where \langle bvexp\rangle\textbf{]}\ \textbf{[}group\ by\ \langle cname\rangle\textbf{]}      \ \textbf{[}having\ \langle bvexp\rangle\textbf{]} 
\end{aligned}
$}
\leftline{$
\begin{aligned}
\ \ \ \ \ \ \textbf{[}order\ by\ \langle cname\rangle      \ [asc|desc]\textbf{]} ]\!] 
\end{aligned}
$}
\leftline{$
\begin{aligned}
\ \ \ \ \ \ \triangleq 
\end{aligned}
$}
\leftline{$
\begin{aligned}
\ \ \ \ \ \ \mathcal{H}[\![from\langle tname_1\rangle]\!]
\end{aligned}
$}
\leftline{$
\begin{aligned}
\ \ \ \ \ \ \ | (\mathcal{H}[\![from\ \langle tname_1 \rangle ]\!],\mathcal{H}[\![from\ \langle tname_2 \rangle ]\!])\diamond \mathcal{C}[\![L,natural/cross\ join\ ]\!] 
\end{aligned}
$}
\leftline{$
\begin{aligned}
\ \ \ \ \ \ \ |(\mathcal{H}[\![from\ \langle tname_1 \rangle ]\!],\mathcal{H}[\![from\ \langle tname_2 \rangle ]\!])\diamond \mathcal{C}[\![L,left/right/inner/full\ join\ ]\!] \diamond \mathcal{H}[\![on\ \langle bvexp\rangle      ]\!]\  
\end{aligned}
$}
\leftline{$
\begin{aligned}
\ \ \ \ \ \ \textbf{[}\diamond\ \mathcal{H}[\![where\ \langle bvexp\rangle      ]\!]\textbf{]}\ 
\end{aligned}
$}
\leftline{$
\begin{aligned}
\ \ \ \ \ \ \textbf{[}\diamond \mathcal{H}[\![group\ by\  \langle cname\rangle      ]\!]\textbf{]}  
\end{aligned}
$}
\leftline{$
\begin{aligned}
\ \ \ \ \ \  \textbf{[}\diamond \mathcal{H}[\![having\ \langle bvexp\rangle      ]\!]\textbf{]} \ 
\end{aligned}
$}
\leftline{$
\begin{aligned}
\ \ \ \ \ \ \diamond\ \mathcal{H}[\![select\ \langle cname_1\rangle     \textbf{[}, \langle cname_2\rangle ...\textbf{]}      ]\!] \textbf{[}\diamond \mathcal{C}[\![\{T_1,\langle sop/af\rangle      \}]\!]\textbf{]}\ 
\end{aligned}
$}
\leftline{$
\begin{aligned}
\ \ \ \ \ \ \textbf{[}\diamond\ \mathcal{H}[\![order\ by\ \langle cname\rangle      \ \textbf{[}asc|desc\textbf{]}]\!]\textbf{]}
\end{aligned}
$}
\leftline{$
\begin{aligned}
104.\ \mathcal{H}[\![\langle queryexp_1\rangle      \ \langle cop\rangle      \ \langle queryexp_2\rangle      ]\!] \triangleq 
(\mathcal{H}[\![\langle queryexp_1\rangle      ]\!],\mathcal{H}[\![\langle queryexp_2\rangle      ]\!])\diamond \mathcal{C}[\![\{[T_1,T_2],\langle cop\rangle      \}]\!]
\end{aligned}
$}
\end{framed}
\caption{The full list of semantic definitions for SQL keywords \texttt{from}, \texttt{select}, \texttt{on}, \texttt{group by}, \texttt{having}, \texttt{order by},  subquery operations, and composite semantics on SQL queries}
\label{fig:fromfullsemantic}
\end{figure*}
We manually analyze the semantics of keywords in the SQL specification and defined the denotational semantics for 138 SQL keywords or features related to query functionality in the SQL specification. 
Although non-trivial manual effort is required to define the semantics, this is a one-off effort. Furthermore, since the semantics of SQL is mostly stable, the defined semantics can be easily maintained as well. 

We use the example in Figure~\ref{fig:prolog-demo} to illustrate the basic components of Prolog. 
The code snippet in line 2 is a fact in Prolog, and it represents a table list containing table \texttt{t} with initialized data.
Lines 4-11 show three rules for the \texttt{SELECT} keyword, corresponding to three types of inputs, i.e., \texttt{NULL} (line 4), constant values (lines 5-8), and lists of columns (lines 9-11).
%
The \texttt{column\_select} functions on lines 8 and 11 correspond to specific column selection operations. 
The first  parameter denotes the columns that the user wishes to select. 
The second parameter contains table metadata, including the table name, column names, and the data items in the table. 
The third parameter is utilized to store the return values. 
Lines 13-15 present a rule for the \texttt{FROM} keyword, where the first parameter \texttt{T} represents the name of the desired table, and the second parameter signifies the output produced by the \texttt{FROM} clause. 
This rule specifically addresses the case of a single table input.
It returns the required table (as shown in line 15) based on the information provided in the facts.  
Lines 17-18 contain Prolog queries, which corresponds to the SQL query \texttt{SELECT t.b FROM t}. 
First, the \texttt{from\_clause} rule (lines 13-15) is first triggered to return table \texttt{t} and then the \texttt{select\_clause} rule (lines 9-11) is activated to return the required column \texttt{t.b}. 
%
In Prolog, the answers to a query can be automatically computed based on rules and facts through a unification algorithm.
As a result, the values in column \texttt{b} are returned and stored in the list \texttt{Filtered} (line 20).

We implement the semantics using Prolog for two main reasons. First, Prolog, like SQL, is a declarative language, which contrasts with imperative languages such as C typically used in RDBMS implementations. This distinction reduces the likelihood of replicating common errors found in traditional RDBMSs. Second, Prolog is intuitive and straightforward to implement, offering built-in support for operations like list manipulation and querying, which align well with the structure of tables and queries in SQL. For these reasons, Prolog is commonly used in existing research to formalize the semantics of various domains~\cite{jiao2020semantic,schumi2022exais}.

\section{Formal semantics of SQL}
\label{sec:semantics}

igure~\ref{fig:wenfa} presents the SQL syntax supported by our system. We implement all keywords and features related to the Data Query Language (DQL), including lexical elements, scalar expressions, query expressions and predicates, as defined in Part 2 of the SQL specification (ISO/IEC 9075-2:2016)~\cite{SQLspec}.

We categorize the denotational semantics of the SQL language into two categories, i.e.,  SQL expression semantics and SQL keyword semantics. 
SQL expressions are basic elements used to construct constraints in queries, while SQL keywords form the main logic of SQL statements.   
%
We adopt the bag semantics of SQL according to the SQL specification, allowing duplicate elements in the result set. 
Moreover, we support the \texttt{null} semantic, which is considered as a special unkown  value in SQL. 
We manually analyze the semantics of keywords in the SQL specification and defined the denotational semantics for 138 SQL keywords or features related to query functionality in
the SQL specification. 
In the following, we first introduce the basic symbols defined for our semantic definitions, then we introduce the semantics of SQL expressions and SQL keywords in detail.

\subsection{Basic symbol definition}
Table~\ref{table:complete set of basic symbol} lists the basic symbols used for semantic definitions in this work.  
%
\texttt{T}, \texttt{$\alpha$}, and \texttt{$\beta$} are used to represent a table, a data record in the table, and an attribute of a table, respectively. $\widehat{\alpha}$ and $\widehat{\beta}$ represent bags of data and bags of attribute, respectively. $\Theta$ represents expressions, including logical expressions, numeric expressions, constant values and function operations defined in SQL. The operations of data projection, data selection, and calculating the multiplicity of data records or attributes are denoted by $\pi$, $\sigma$, and $\xi$, respectively. The operations of joining two data records and performing the Cartesian product are represented by $\bowtie$ and $\times$, respectively. 

Furthermore, this work uses $[a,...,a]_n$ to denote a list of n occurrences of $a$ (which can be a data record or a constant), $max(a_1,a_2)$ and $min(a_1,a_2)$ to represent the operation of finding the maximum or minimum value between $a_1$ and $a_2$. The functions $num2str(n)$ and $str2num(s)$ are used for converting between numeric and string types, $type(a)$ returns the type of data record $a$.

\subsection{Semantics of Expressions}
Fig~\ref{fig:boolfullsemantic} and ~\ref{fig:numstrfullsemantic} show our semantic definition for  the expressions in SQL. 
In SQL, expressions are essential structural elements used to construct queries and compute return values. The complexity of an expression can vary, ranging from simple numeric or string values to complex combinations of functions, operators, and subqueries. 
We categorize the semantics of SQL expressions into boolean, numeric, and string value expression according to the specific return value type. 

\begin{table}[t]   
\begin{center}   
\caption{Symbol notations}
\label{table:complete set of basic symbol} 
\small
\begin{tabular}{rl}   
\toprule   
Symbol & Description \\      
\midrule
\texttt{T} & A table instance \\
$ \alpha$ & A data record in a table  \\ 
$ \widehat{\alpha}$ & Bags of data records  \\
$ \beta$ & An attribute in a table  \\
$ \widehat{\beta}$ & Bags of attributes  \\
$ \Theta $ & An expression\\
 \texttt{T}.$\beta_i$ & The i$_{th}$ attribute in \texttt{T}  \\
\texttt{T}.$\overline{\beta}$ & All attributes in  \texttt{T}\\
$ \pi_{\beta}(\texttt{T})$ & Data projection on attribute $\beta$  \\
$\sigma_{\Theta}(\texttt{T})$ & Data selection on expression $\Theta$  \\
$ \xi_{\alpha}(\texttt{T})$ & The multiplicity of data record $\alpha$ in table \texttt{T}\\
$ \xi_{\beta}(\texttt{T})$ & The multiplicity of 
attribute $\beta$ in table \texttt{T} \\
$ [a,...,a]_n$ & A list of element a with length n   \\
$\bowtie$ & Data join operation \\
$\times$ & Cartesian product operation   \\
$max(a_1,a_2)$ &The maximum value in $a_1 $and $a_2 $    \\
$min(a_1,a_2)$ &The minimum value in $a_1 $and $a_2 $    \\
$num2str(n)$ & Converting the numeric type n to a string   \\
$str2num(s)$ & Converting the string type s to a number  \\
$type(a)$ &The data type of a    \\
$ [a,...,a]_n$ & A data list of $a$ of length n   \\
\bottomrule  
\end{tabular}   
\end{center}   
\end{table}

\noindent\textbf{Definition 1 (\text{Boolean value expression ($\mathcal{B}:E_B \mapsto \mathbb{B}$)})} Fig~\ref{fig:boolfullsemantic} (Semantic rules 1-34) lists the semantics of SQL expressions, including logical expressions, comparison expressions, IS expressions, BETWEEN expressions, IN expressions, NULL expressions, EXISTS expressions, and relevant functionalities that perform implicit type conversions on other types of expressions, that return boolean values. The function $\mathcal{B}$ maps from the domain of boolean value expressions $E_B$ to the domain of Boolean values $\mathbb{B}$.  

\noindent\textbf{Definition 2 (\text{Numeric value expression ($\mathcal{N}:E_N \mapsto \mathbb{N}$)})}  Fig~\ref{fig:numstrfullsemantic} (semantics rules 35-59) show the semantics of 
SQL expressions, including arithmetic expressions, numeric functions, and relevant functionalities that perform implicit type conversions on the results of other types of expressions, that return numeric values. 
The function $\mathcal{N}$ maps from the domain of numeric value expressions  $E_N$  to the domain of numeric values $\mathbb{N}$.  

\textbf{Definition 3 (\text{String value expression ($\mathcal{S}:E_S \mapsto \mathbb{S}$)})} Fig~\ref{fig:numstrfullsemantic} (Semantic rules 60-75) shows the semantic rules of SQL expressions, including include string concatenation expressions, string functions, and relevant functionalities that perform implicit type conversions on the results of other types of expressions, that return string values. 
The function $\mathcal{S}$ maps from the domain of string value expression $E_S$ to the domian of string values $\mathbb{S}$. 


\noindent\subsection{Semantics of SQL keywords}
The semantics of SQL keywords can be classified into two categories. The first category, including join operations, set operations, filtering operations, and aggregate functions,  involves functionalities that directly operate on a list of tables and return a new table.  The second category, encompassing the semantics of keywords such as FROM and SELECT, as well as their composite semantics, involves functionalities that operate on query expressions, and return a tuple (A, T), where A is the set of all tables and T is the resulting table.

\noindent\textbf{Definition 4 (\text{Keyword operation ($\mathcal{C}:\{L,OP\} \mapsto T$)})} Fig~\ref{fig:joinfullsemantic} shows the semantic definition of four SQL keywords, i.e., JOIN operations (semantic rules 76-80), set operations (semantic rules 81-83), filter operations (semantic rules 84-85), and aggregate operations (semantic rules 86-90). These operations take a list of table instances $L$ and an operation type $OP$ as input and produce a new table instance $T$ as output. 

\textbf{Definition 5 (\text{Keyword operation ($\mathcal{H}:S \mapsto (A,T)$)})} Fig~\ref{fig:fromfullsemantic} shows the semantics of the second category of SQL keywords (semantic rules 91-101), including keywords such as FROM, WHERE, ON, SELECT, GROUP BY, HAVING, ORDER BY, and their combinations in SQL statements. 
The function $\mathcal{H}$ is a mapping from the domain of SQL statements $S$ to the domain of tuples $(A, T)$, where $A$ represents the set of tables that are relevant or affected during the execution process. 
$T$ represents the table obtained after the execution. 

\textbf{Definition 6 (\text{Composite operation ($\mathcal{H}:S \mapsto (A,T)$)})} Based on the semantics of these keywords and functions, we define the composite semantics of SQL language to represent the semantics of a SQL statement. The input of the composite semantics is a SQL statement, and the output is a tuple $(A, T)$. As shown in Fig~\ref{fig:fromfullsemantic}, we first introduced the combination rules of different types of keyword semantics (semantic rule 102), and then we introduce the composite semantics based on those rules (semantic rules 103-104).

\section{Conformance Testing}

\subsection{Overview of our approach}

\begin{figure}[t]
\centerline{\includegraphics[width=0.50\textwidth,height=0.30\textwidth]{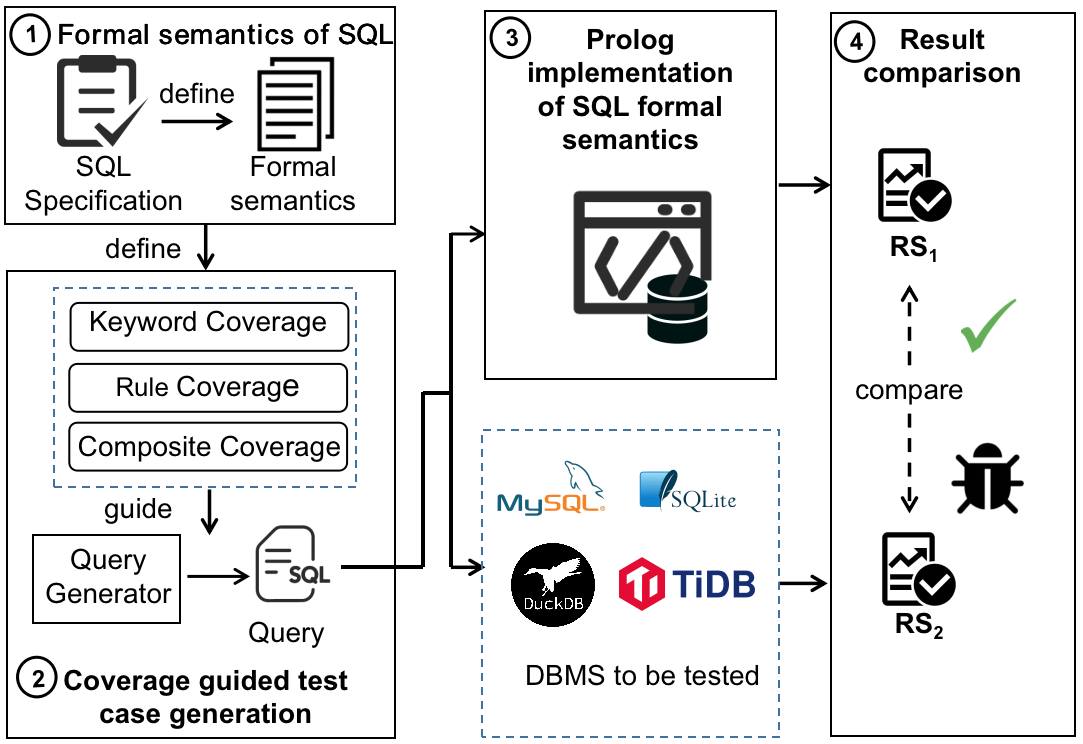}}
\caption{Overview of \tool{} 
}
\label{fig:overview}
\end{figure}

Figure~\ref{fig:overview} illustrates the overview of our conformance testing approach, which consists of four components. 
Initially, we define the formal semantics of SQL, as detailed in Section~\ref{sec:semantics}.
Next, we implement the SQL formal semantics in Prolog, which serves as an oracle for conformance testing. 
The third component is dedicated to test case generation. 
Here, we enhance the syntax-guided generation method, SQLancer~\cite{sqlancer}, with coverage-guided test case generation. 
This enhancement is based on three coverage criteria, i.e., keyword coverage, rule coverage, and composite rule coverage, which we proposed based on the formal semantics we defined. 
The final component is dedicated to query results comparison, wherein the query results of the tested RDBMS and our Prolog implementation are compared to identify conformance issues.

\textbf{Conformance issues} in our work refer to two types of issues, i.e., bugs or inconsistencies. Both issues are due to violating the SQL semantics defined in the SQL specification, or unclear or missing descriptions in the SQL specification. \textbf{Bugs} are defects that are confirmed by developers. \textbf{Inconsistencies} refer to inconsistent result produced by the tested RDBMS and \tool{}. Inconsistencies are also confirmed by developers, yet they perceive them as deliberate design choices rather than bugs. We report these inconsistencies because various RDBMSs make differing design decisions, leading to varied query results that may potentially perplex users. This also underscore the importance of a comprehensively documented SQL specification. 


\subsection{Coverage guided test case generation}

The state-of-the-art practice in test case generation involves randomly generating SQL queries guided by the syntax of SQL, 
among which SQLancer~\cite{sqlancer} stands out as one of the most effective tools of this kind. 
It 
is tailored to the syntactical structures of various RDBMSs. 
SQLancer considers database objects, such as tables, views, and indexes, as well as keywords and functionalities within query statements. 
Throughout the generation process, SQLancer maintains a set of keywords and functionalities, from which it randomly selects keywords to incorporate into the test cases, subject to syntactic rules of SQL (so that they remain syntactically valid).
However, this generation process is entirely random and does not consider coverage of the SQL semantics.
As a result, certain aspects of the semantics may never be tested.

To address this problem, we propose three coverage criteria, i.e., keyword coverage, rule coverage, and composite rule coverage, based on the formal semantics we defined.
Each coverage criterion defines coverage at a different granularity level. 
Keyword coverage simply assesses whether each individual keyword in the SQL specification is covered. 
Rule coverage goes a step further by calculating whether each semantic rule, which accommodates different inputs for a SQL keyword, is covered. 
Composite rule coverage considers combinations of semantic rules that can form a valid SQL query.

\noindent\textbf{Keyword coverage.} 
Formula~(\ref{eq:keywordcov}) presents the formula for calculating keyword coverage,  
where $N_{tk}$ represents the number of keywords that are covered by the generated test queries and $N_K$ the total number of keywords defined in our semantics. 
\begin{equation}\label{eq:keywordcov}
Cov_k = \frac{N_{tk}}{N_K}
\end{equation}


To calculate keyword coverage, we count the total non-repeated occurrences of keywords in all test queries
as the number of $N_{tk}$. 
In our case, $N_K$, which represents the total count of SQL keywords defined by our semantics, is 138. 
 



\noindent\textbf{Rule coverage.}
The semantics of each SQL keyword, as defined in Prolog, is often captured using multiple rules. 
Formula~(\ref{eq:rulecov}) shows the formula, for measuring the percentage of rules covered by the generated test queries ($N_{tr}$) in relation to the total number of rules ($N_R$). 
The total number of rules, $N_R$, is the sum of all rules for all keywords we have defined in our semantics. 
$N_{tr}$ is calculated by enumerating all keywords in the test suite and identifying the distinct semantic rules that are triggered. 

\begin{equation}
    \label{eq:rulecov}
    Cov_r=\frac{N_{tr}}{N_R}
\end{equation}


Note that we retain all duplicate keywords when collecting the semantic rules, as SQL queries with the same keyword may trigger different rules. 
We only remove duplicate semantic rules after collecting the complete set.
For instance, as shown in Figure~\ref{fig:prolog-demo}, the semantics for the keyword \texttt{SELECT} include three rules, i.e., line 4 for \texttt{SELECT NULL}, lines 5-8 for the case of a constant value input, and lines 9-11 for the case of column selection in the table. 
The SQL query \texttt{SELECT t.b FROM t} covers the third \texttt{SELECT} rule and the first \texttt{FROM} rule, resulting a rule coverage of 2/420, where 420 is the total number of different rules defined in our semantics.

\begin{algorithm}[t]
\caption{Calculating the number of composite rules}
\label{alg:total-composite-rules}
\small
 \SetAlgoLined
\SetKwFunction{TCut}{TraverseGrammar} 
\SetKwFunction{MCut}{CRNumberCalculation}
\SetKwInOut{Input}{Input}
\SetKwInOut{Output}{Output}
    \Input{
        $grammar$:  SQL grammar \\
       }
    \Output{
        $N_{CR}$: the number of composite rules\\
    }
    Initialize $N_{CR}$=0, $path$=empty \\
    \SetKwProg{Fn}{Function}{:}{}
    \Fn{\MCut{$grammar$}}{
        \textit{TraverseGrammar} ($grammar.headNode$, $path$)\\
        \Return{$N_{CR}$}
    }
    \SetKwProg{Fn}{Function}{:}{}
    \Fn{\TCut{$node$, $path$}}{
        \If{$node$ is Non-terminal}{
            \If{$node$ is SQLKeyword}{
                $path.\text{add}(node)$ \\
            }
            \If{$node.children$ is not Null}{
                \ForEach{$child$ in $node.children$}{
                    \textit{TraverseGrammar}($child$, $path$)
                }
            }        
        }    
        \Else{
            $N_{pcr}$ = 1 \\
            \ForEach{$Keyword$ in $path$}{
                $N_{pcr}$ =  $N_{pcr}$ * $Keyword.ruleNumber$
            }
            $N_{CR}$ = $N_{CR}$ + $N_{pcr}$ 
        }
        $path.\text{remove}(node)$
    }   
\end{algorithm}

\noindent\textbf{Composite rule coverage.}
%
We further introduce a more fine-grained coverage criterion named composite rule coverage, which takes into account the combination of semantic rules triggered by a SQL statement, offering a more comprehensive assessment of the test coverage for SQL statements. 
We calculate the composite rule coverage using Formula~\ref{eq:compositerule}. 
The numerator ($N_{tcr}$) represents the number of composite rules covered by the test queries, while the denominator ($N_{CR}$) is the total number of all composite rules.

\begin{equation}
\label{eq:compositerule}
    Cov_{cr}=\frac{N_{tcr}}{N_{CR}}
\end{equation}

Algorithm~\ref{alg:total-composite-rules} outlines the process of calculating the total number of composite rules. 
The input to the algorithm is the SQL grammar, and its output is the total number of composite rules ($N_{CR}$).
The \textit{path} variable declared in line 1 records each traversed paths, which are SQL keyword sequences according to grammar rules. 
Our algorithm conducts a depth-first traversal following the SQL grammar rules by recursively calling the \texttt{TraverseGrammar} function from the headnode of the grammar (lines 2-4). 
\texttt{TraverseGrammar} initially checks whether a node in the grammar is a non-terminal node. 
If it is both a non-terminal node and a SQL keyword, we include it in the path. 
A depth-first search is then performed by recursively invoking function \texttt{CRNumberCalculation} (lines 11-13). 
When a leaf node is incurred, we calculate the composite rules of the path by multiplying the number of rules of each keyword, i.e., node in the path (lines 18-20). 
The composite rules for this path are then added to the total number of composite rules (line 21). 
Upon completion of the traversal, the current value of $N_{CR}$ is returned.

Take the statement \texttt{SELECT t.b FROM t} and the Prolog rules in Figure~\ref{fig:prolog-demo} as an example. 
To identify the composite rules covered by this query, we enumerate the rules of the \texttt{SELECT} and \texttt{FROM} keywords and match the rules based on the parameter or input of that keyword. 
In this case, the \texttt{SELECT} rule that takes a column list as input and the \texttt{FROM} rule that takes one table as input are matched to form the composite rule covered by the given query.

The time complexity of Algorithm~\ref{alg:executeSQL} is O(n) with n being the number of operators in the given SQL query.


\begin{algorithm}[t]
\caption{Coverage guided query generation}
\label{alg:rebuild-data-dep}
\small
 \SetAlgoLined
\SetKwFunction{TCut}{SQLGeneration} 
\SetKwFunction{MCut}{CalculateCoverage}
\SetKwInOut{Input}{Input}
\SetKwInOut{Output}{Output}
    \SetKwProg{Fn}{Function}{:}{}
    \Fn{\TCut{}}{
        Initialize $coverage$=0, $coveredSet$\\
        Initialize $queryPool$=GenerateQuery()\\
        \While{TRUE} {
            \While{$coverage$ does not increase}{
                $queryInit$ = \textit{GetQueryFromPool}() \\
                $query$ = \textit{MutateQuery}() \\
            \textit{CalculateCoverage}($query$) \\
            }
            \textit{AddQueryIntoPool}($query$)\\
            \textit{ExecuteQuery}($query$)\\
            \If{timeout}{
                break
            }
        }        
    }
    \SetKwProg{Fn}{Function}{:}{}
    \Fn{\MCut{$query$}}{
        \If{$query.pattern$ not in $coveredSet$}{
            \textit{UpdateCoverage}($coverage$)\\
            ADD $query.pattern$ To $coveredSet$ \\
        }
    }
   
\end{algorithm}

\noindent\textbf{Coverage-guided query generation.} 
\begin{table*}
\setlength{\abovecaptionskip}{0cm} 
\centering
  \caption{Mutation Rules, with Colored \sout{\textcolor{violet}{Deletion}} and \textcolor{violet}{Addition}}
    \label{tab:MR}
 \resizebox{\textwidth}{!}{
\begin{tabular}{clll}
    \toprule
    \multirow{1}{*}{ID}&\multirow{1}{*}{Type}&\multirow{1}{*}{Transformation}&\multirow{1}{*}{Example Query} \\
    \midrule
     01&\multirow{6}{*}{Keyword-level}&Replace operators&SELECT * FROM T WHERE T.a \sout{\textcolor{violet}{AND}} \textcolor{violet}{OR} T.b\\
     02&&Replace keywords&SELECT * FROM T \sout{\textcolor{violet}{ORDER BY}} \textcolor{violet}{GROUP BY} T.a\\
     03&&Add operators&SELECT * FROM T WHERE (T.a AND T.b) \textcolor{violet}{IS NOT TRUE}\\
     04&&Add keywords&SELECT * FROM T WHERE T.a = CAST(T.b as string) \textcolor{violet}{ORDER BY T.b ASC}\\
     05&&Delete operators&SELECT * FROM T WHERE T.a  \sout{\textcolor{violet}{AND T.b}} \\
     06&&Delete keywords&SELECT * FROM T  \sout{\textcolor{violet}{WHERE EXP(T.a) $>=$ T.b}} \\
     \midrule
    07&\multirow{4}{*}{Rule-level}&Convert constants to column references& SELECT * FROM T WHERE T.a = MOD(\sout{\textcolor{violet}{1}} \textcolor{violet}{T.b},1)\\
    08&&Convert parameter data types& SELECT * FROM T WHERE T.a = POSITION(\sout{\textcolor{violet}{1}} \textcolor{violet}{'a'},1)\\
    09&&Add parameters& SELECT * FROM T WHERE T.a = MOD(T.b,1) IS NOT TRUE \textcolor{violet}{AND ABS(1)}\\
    10&&Delete parameters& SELECT * FROM T WHERE T.a = (T.b $>$ FLOOR(T.c) \sout{\textcolor{violet}{XOR CEILING(1.5)}}) \\
    \midrule
    11&\multirow{3}{*}{Subquery-level}&Replace subqueries&SELECT * FROM T \sout{\textcolor{violet}{WHERE T.a = 1 AND T.b
    IS FALSE}} \textcolor{violet}{WHERE EXISTS SELECT T.c WHERE T.b =1}\\
     12&&Add subqueries&SELECT * FROM T \textcolor{violet}{WHERE T.a IN (1,2) XOR T.b = exp(3)}\\
     13&&Delete subqueries&SELECT * FROM T  WHERE T.a $>$ (T.b IS NOT UNKNOWN) \sout{\textcolor{violet}{GROUP BY T.a HAVING LN(4)}} \\
  \bottomrule
\end{tabular}
}
\end{table*}

Algorithm~\ref{alg:rebuild-data-dep} describes the process of query generation guided by coverage. 
We set the initial coverage to be 0 and the declare covered set (\textit{coveredSet}), which records the covered keywords, rules, or combined rules. 
We adopt SQLancer to randomly generate a large number of SQL statements (line 2), which serves as the seed pool of our query generation algorithm. 
The algorithm begins with randomly selecting an SQL statement from the seed pool (line 6), and mutates the query based on the mutation rules we proposed. 
Then we calculate coverage of the mutated query (line 8).  We keep this mutation process (line 5-9) until the mutated query increases the overall coverage.  
Then the mutated query is added into the seed pool for future test case generation.
The mutated query is executed to explore potential inconsistencies (line 11). 
The process terminates upon timeout.   
Function \texttt{CalculateCoverage} calculates the coverage of the given query. It first checks whether the given query's signature according to our definition of coverages (i.e., keyword, rule, or composite rule) is already recorded in the covered set. 
If it is not in the covered set, it indicates that this query increases the coverage. 
We then update the coverage (line 18) and add the pattern of this query to the covered set (line 19).

The time complexity of Algorithm~\ref{alg:rebuild-data-dep} is O(nlogn), with n being the number of all possible rules for a particular coverage criterion. 
Our goal is to generate a set of test cases that collectively cover all semantic rules. The test case generation algorithm operates by randomly generating a test case and retaining it only if it covers a previously uncovered rule; otherwise, it is discarded. This process is analogous to the Coupon Collector’s problem~\cite{CouponCollector}, which estimates the time required to collect n distinct coupons through random sampling. Similarly, our random generation process achieves full coverage with high probability at a time complexity of O(nlogn). In our work, the number n is 138 for keyword coverage, 556 for semantic rule coverage and 19 million for composite rule coverage.

Table~\ref{tab:MR} lists the mutation rule examples on SQL statements we proposed. We categorize the mutation rules into three classes, i.e., keyword-level mutation rules, parameter-level mutation rules and subquery-level mutation rules. These mutation rules effectively enhance the keyword coverage, rule coverage, and combination rule coverage. 
In particular, keyword-level mutation rules and subquery-level mutation rules improve keyword coverage, parameter-level mutation rules improve rule coverage, all three types of mutation rules used together improve composite rule coverage. 
To ensure the validity of the mutated queries, we employ the SQL parser JSQLParser~\cite{JSQLParser} during the mutation process to verify the syntactic validity of each mutated statement and discard the ones with syntax errors. 
Semantic checks, including the table references, column references, and data types, are performed to ensure that the correctness of the mutated queries. 

\subsection{Prolog implementation of SQL formal semantics}

\begin{algorithm}[t]
\caption{Executing a SQL query}
\label{alg:executeSQL}
\small
 \SetAlgoLined
\SetKwFunction{MCut}{ExecuteQuery}
\SetKwFunction{NCut}{ExecuteKeyword}
\SetKwInOut{Input}{Input}
\SetKwInOut{Output}{Output}
    \Input{
        $sql$: the SQL query to be executed  \\
       }
      \Output{
        $result$: the execution result of the query \\
      }
$ast$ = \textit{ParseSQL} ($sql$)\\

 \SetKwProg{Fn}{Function}{:}{}
    \Fn{\MCut{$ast.root$}}{
            $keywordList$ = \textit{Sort}($ast.root.children$)\\
            \ForEach{$keyword$ in $keywordList$} {
                $result$ = \textit{ExecuteKeyword}($keyword, result$)    
            }
    }
    \SetKwProg{Fn}{Function}{:}{}
    \Fn{\NCut{$keywordNode, result$}}{
        \ForEach{$child$ in $keywordNode.children$} {
            \If {$child$ is \textit{query}}{
                $result$ = \textit{ExecuteQuery}($child$)
            }  
            \If{child is leaf}{
            $result$=\textit{ExecuteRule}($keywordNode, result$)\\
            \textbf{return} $result$ \\
        }
            $result$ = \textit{ExecuteKeyword}($child, result$)
        }
    }
\end{algorithm}

We have implemented the formal semantics of SQL defined in Figure 4-7 in Prolog. 
The semantics of each keyword are implemented as a set of rules, as illustrated in Figure~\ref{fig:prolog-demo}. 
We then implement the compound semantics in Algorithm~\ref{alg:executeSQL}, which outlines the process of executing a SQL query in Prolog. 
The input to this algorithm is a SQL query, and its output is the result of executing this SQL query. 
Algorithm~\ref{alg:executeSQL} initially parses the SQL statement into an Abstract Syntax Tree (AST) (line 1). 
The \texttt{ExecuteQuery} function then traverses the tree from the root node, sorting the children of the root node according to the keyword execution order (line 3). 
Then the semantic rules of the keywords are executed in order with the \texttt{ExecuteKeyword} function (lines 4-6).  
If a subquery is encountered (lines 9-11), the \texttt{ExecuteQuery} function is recursively called to initiate the sorting procedure. 
In other cases, \texttt{ExecuteKeyword} recursively calls itself (line 16) until a leaf node is reached, which invokes the corresponding keyword semantic rule execution (lines 12-14).


The sorting of keywords solves the critical issue of ensuring the correct execution order of the semantics for each keyword in the query. 
Note that the SQL specification does not explicitly indicate the execution order of all keywords in a query, yet we can imply the execution order based on the semantic of each individual keyword.  
We also check the implementation of current mainstream databases, including MySQL, PostgreSQL, TiDB, SQLite and DuckDB, and confirm that they enforce the same execution order of SQL keywords, which is consistent with our understanding of keyword execution order, i.e., \texttt{JOIN}, \texttt{FROM}, \texttt{WHERE}, \texttt{GROUP BY}, \texttt{Aggregate functions}, \texttt{HAVING}, \texttt{SELECT}, \texttt{ORDER BY}, based on their semantics.

\begin {comment}\tcl{
In the SQL specification, there are a total of 47 keywords whose semantics are not explicitly described, among which 4 are duplicated, e.g., AND and $\&\&$, OR and $||$, LCASE and LOWER, UCASE and UPPER. The remaining 43 keywords include 6 bitwise operators, 25 are string functions, and 16 are numeric functions. Specifically, there are four keywords that have semantic duplications, such as . Regarding the remaining 43 keywords lacking descriptions, taking bitwise operators as an example, the SQL specification (\textit{Part 2 Foundation, Language Opportunities}) states: ``The SQL standard is missing operators on binary data types (BINARY, VARBINARY, BLOB) that allow users to bitwise manipulate values."
For these keywords, we referred to the implementation documentation of current mainstream database management systems to guide the process of implementing their semantics. In cases where there were inconsistencies among different database implementations, we chose to adopt the approach used by the majority of databases.
}
\end{comment}

In the SQL specification, there are a total of 47 keywords whose semantics are not explicitly described, among which 4 are duplicated, e.g., AND and $\&\&$, OR and $||$, LCASE and LOWER, UCASE and UPPER. The remaining 43 keywords include 4 bitwise operators, 23  string functions, and 16 numeric functions. Taking bitwise operators as an example, the SQL specification(\textit{Part 2 Foundation, Language Opportunities}) states: "The SQL standard is missing operators on binary data types (BINARY, VARBINARY, BLOB) that allow users to bitwise manipulate values."
For these keywords, we referred to the implementation documentation of current mainstream database management systems in our Prolog implementation of semantics. In cases where there were inconsistencies among different database implementations, we chose to adopt the approach used by the majority of databases.

Take the SQL query \texttt{SELECT * FROM T WHERE T.a=(SELECT 1 FROM T)} as an example. 
This query contains a subquery  \texttt{SELECT 1 FROM T}. 
Initially, we parse this SQL statement into an AST and sort the three children nodes of the root node according to the keyword execution order of \texttt{FROM}, \texttt{WHERE}, \texttt{SELECT}. 
Taking the \texttt{FROM} keyword as an example, it has a single child node, which is the table name \texttt{T}. 
The rule for \texttt{FROM} that requires a table as input is then triggered for execution (lines 12-14). 
This information is subsequently relayed to the \texttt{WHERE} clause. 
During the execution of the \texttt{WHERE} clause, we encounter the subquery \texttt{SELECT 1 FROM T}, where we recursively call the \texttt{ExecuteQuery} function to process the subquery.

\noindent\textbf{Correctness of \tool{}. }
As mentioned in Section~\ref{sec:preliminary}, Prolog, being declarative, is naturally suited to specify denotational semantics we defined.  
Taking the `select' keyword as an example, rule 19 of Figure~\ref{fig:joinfullsemantic} shows the formal semantics of `select' and line 9-11 in Figure~\ref{fig:prolog-demo} shows the corresponding Prolog implementation. The formal semantics defines the select keyword as a column reference using the projection operation $\pi$ to select columns from a table \texttt{T}. This maps directly to the select\_clause function in the Prolog implementation, which checks for column references and extracts the relevant columns from Table \texttt{Tb}. Prolog's rule-based structure ensures a one-to-one correspondence with the formal semantics, minimizing implementation errors and ensuring adherence to the SQL specification. 
Moreover, we have conducted comprehensive testing and code review following the software engineering standard procedure, covering all the semantic rules we've implemented. 
We also conducted thorough experiments with 6 different RDBMS systems, including MySQL, PostgreSQL, TiDB, SQLite, DuckDB and OceanBase, validating the correctness our implementation using 18 millions of test cases.

\subsection{Result comparison}

The final part of our method involves result comparison. 
This process entails comparing the query results from the tested RDBMS with those returned by \tool{}. 
We first compare the number of records in the query results and identify an inconsistency if the numbers differ. 
If the numbers are identical, we proceed to compare the data records in the results. 
In particular, we scan both sets of query results and remove identical data pairs. 
An inconsistency is reported if either result set is not empty after removing all matching pairs. 

For some of the SQL features, such as arithmetic operators, aggregate functions, and numerical functions, different RDBMSs may incur different implementation choices on floating point precision, which could result in false alarms in our result comparison step. To mitigate those false alarms, we impose restrictions on the return results of SQL statements that may involve floating-point outcomes during test case generation, and enforce the execution results  to retain two decimal places. Meanwhile, we impose the same restrictions in our implementation of SQL semantics in Prolog to avoid potential false alarms in result comparison.

 To enhance result explainability reported by \tool{},  
we log the inconsistencies and provide the semantic rule we implemented as explanations of the inconsistency. Moreover, for those under-specified keywords like \texttt{IN}, we provide multiple Prolog implementations based on popular RDBMSs, e.g., MySQL and PostgreSQL, allowing users to configure the desired semantics.

\subsection{Discussion on Extensibility} 
\label{sec:discussion}
In this paper, we define and implement the denotational semantics concerning the SQL Data Query Language (DQL) commands. The other types of SQL commands, including DDL, SML, and DCL can be easily supported by extending our semantics. 
For the semantics of transactions and concurrency, we formalize single transactions by executing SQL queries sequentially in real-time order. For concurrent transactions, the semantics should define all valid schedules. Formally, the semantics of two concurrent transactions T1 and T2 can be defined as:  
$
T1||T2\triangleq \{Q_1 || Q_2, Q_1 \in T1, Q_2\in T2 \land RTConstraint(T1) \land RTConstraint(T2) \land !IsolationConstraint\}   
$ 
The symbol $||$ represents the concurrent execution of two transactions or SQL queries, \textit{RTConstraint(T)} formalizes the realtime order constraints of SQL queries in T, and \textit{IsolationConstraint} formalizes the schedule constraints associated with a particular isolation level.  
Since the SQL specification~\cite{SQLspec} only provide the anomaly phenomena, which can be formalized as specific schedule templates among transactions, to be avoided in each isolation level, we need to exclude those invalid schedules in our semantics. 

Taking dirty read as an example, any schedule that contains the sequence of \texttt{T1.w(x), T2.r(x)} should be avoided as T2 has read an uncommitted write by T1, and this potentially lead to dirty read if T1 aborts. Then this pattern can be added into the \textit{IsolationConstraint} to filter out schedules containing this pattern. This schedule constraint is associated with all isolation levels as they all forbid dirty read.  
We can generate test cases that contain schedules of the phenomena to be avoided and inspect on the logs of the tested RDBMSs to check whether their implementations contain behaviors of those phenomena. 

\section{Evaluation}

\begin{table}[t]
\footnotesize
\begin{center}   
\caption{Statistics of the target RDBMSs
}  
\label{tab:statistics} 
\scalebox{1.0}{\begin{tabular}{cccccccccccc} 
\toprule
\multicolumn{2}{c}{\multirow{2}{*}{DBMS}}& \multicolumn{2}{c}{Popularity rank}& \multicolumn{1}{c}{\multirow{2}{*}{\tabincell{c}{LOC}}}& \multicolumn{1}{c}{\multirow{2}{*}{\tabincell{c}{First \\release}}}&\multicolumn{1}{c}{\multirow{2}{*}{\tabincell{c}{Tested \\version}}}\\
\cmidrule{3-4}
\multicolumn{2}{c}{}&\tabincell{c}{DB-engines}&\tabincell{c}{GitHub stars}&{}&{}&{}\\  
\midrule  
\multicolumn{2}{c}{MySQL}&2&5.0K& 380K&1995 & 8.0.29\\
\multicolumn{2}
{c}{TiDB}&118&23.1K& 800K&2017 & 6.6.0\\
\multicolumn{2}{c}{SQLite}& 9& 1.5K& 300K& 2000 & 3.39.0\\
\multicolumn{2}{c}{DuckDB}&-&0.5K& 59K&2018 & 0.7.0\\
\multicolumn{2}{c}{PostgreSQL}&4&14.9K& 125K&1996 & 16.1\\
\multicolumn{2}{c}{OceanBase}&119&8.3K& 300K&2015 & 4.3.2\\
\bottomrule
\end{tabular}}
\end{center}   
\end{table}

In this section, we evaluate the effectiveness of \tool{}. 
Specifically, our experiment is designed to answer the following four research questions.

\noindent\textbf{RQ1:} Can \tool{} detect conformance issues, i.e., bugs and inconsistencies, in relational DBMS systems?

\noindent\textbf{RQ2:} Are the three coverage criteria effective in guiding the generation of queries for uncovering bugs and inconsistencies in RDBMSs?

\noindent\textbf{RQ3:} How does \tool{} perform compared to state-of-the-art RDBMS testing approaches?

\noindent\textbf{RQ4:} Case studies of the bugs and inconsistencies detected by \tool{}. 

\subsection{Experiment setup}

We conducted all experiments on a server with two Intel(R) Xeon(R) Platinum 8260 CPUs at 2.30 GHz and 502 GB of memory, running Ubuntu 18.04.6 LTS. 
The SQL formal semantics were implemented in Prolog, while the SQL query generation program was developed in Java. 
We ran the experiments using Java version 11.0.15.1.



\noindent\textbf{Target RDBMS.} 

We selected six popular and widely used RDBMSs, each offering a range of distinct features and application scenarios, to demonstrate the effectiveness of our approach. 
The statistics of these RDBMSs, as obtained from their open-source repositories, are shown in Table~\ref{tab:statistics}. 
It is important to note that we used the latest release of each RDBMS, which has been extensively tested by existing approaches~\cite{rigger2020detecting, rigger2020finding}. 
MySQL~\cite{mysql} and PostgreSQL~\cite{psql} are the two most popular open-source database management systems. 
SQLite~\cite{sqlite} and DuckDB~\cite{duckdb} are both embedded DBMSs, running within the process of other applications. 
TiDB~\cite{tidb} and OceanBase~\cite{oceanbase} are popular distributed RDBMSs.

\noindent\textbf{Compared baselines.} 
We compared \tool{} with TLP~\cite{rigger2020finding} and NoREC~\cite{rigger2020detecting}, which are state-of-the-art  metamorphic testing methods for testing RDBMS. NoREC constructs two semantically equivalent queries, one triggers the optimization and the other does not, executes the queries and compare the results. 
TLP, on the other hand, partitions the conditional expression of the original query into three segments, corresponding to the three possible results, i.e., TRUE, FALSE, and NULL, of the conditional expression. 
It then compares the union of the result sets from executing the three queries with the three segments each, with the result set of the original statement, expecting them to be identical. 
Both approaches have demonstrated effectiveness in RDBMS bug detection.
Both NoREC and TLP are implemented in SQLancer~\cite{sqlancer} and SQLRight~\cite{liang2022detecting}. SQLancer adopts a generative approach for query generation and SQLRight adopts a mutation-based approach for generating queries. 
Therefore, in our experiment, we have four combined settings (concerning the oracle and query generation method) for the compared baselines, i.e., NoREC (SQLancer), NoREC (SQLRight), TLP (SQLancer) and TLP (SQLRight).

\begin{table}[t]
\centering 
\scriptsize
\caption{Bugs and inconsistencies detected by \tool{} 
} 
\label{tab:bugs} 
\begin{tabular}{cccccc} 
\toprule 
SN & ID & Target & Type & Reason & Status\\
\midrule
1 & 109146 & MySQL & Bug &missing spec &duplicate\\
2 & 109837 & MySQL & Bug &missing spec&comfirmed\\
3 & 109842 & MySQL & Bug &missing spec&comfirmed\\
4 & 109149 & MySQL & Bug &missing spec&comfirmed\\
5 & 110438 & MySQL & Bug &violate spec&comfirmed\\
6 & 109147 & MySQL & Inconsistency &missing spec&comfirmed\\
7 & 109148 & MySQL & Inconsistency &unclear spec&comfirmed\\
8 & 109836 & MySQL & Inconsistency &missing spec&comfirmed\\
9 & 109845 & MySQL & Inconsistency &missing spec&comfirmed\\
10 & 110439 & MySQL & Inconsistency &violate spec&comfirmed\\
11 & 110346 & MySQL & Inconsistency &violate spec&comfirmed\\
12 & 109962 & MySQL & Inconsistency &missing spec&comfirmed\\
13 & 110711 & MySQL & Inconsistency &missing spec&comfirmed\\
\midrule
14 & 40996 & TiDB & Bug &missing spec&comfirmed\\
15 & 40995 & TiDB & Bug &missing spec&comfirmed\\
16 & 39260 & TiDB & Bug &missing spec&comfirmed\\
17 & 39259 & TiDB & Bug &missing spec&comfirmed\\
18 & 39258 & TiDB & Bug &unclear spec&comfirmed\\
19 & 42375 & TiDB & Bug &violate spec&comfirmed\\
20 & 42376 & TiDB & Bug &violate spec&comfirmed\\
21 & 42378 & TiDB & Bug &violate spec&comfirmed\\
22 & 42379 & TiDB & Bug &violate spec&comfirmed\\
23 & 42377 & TiDB & Bug &violate spec&comfirmed\\
24 & 42773 & TiDB & Inconsistency &missing spec&comfirmed\\
25 & 40995 & TiDB & Inconsistency &missing spec&comfirmed\\
\midrule
26 & 7e03a4420a  & SQLite & Bug &missing spec&comfirmed\\
27 & 3f085531bf  & SQLite & Bug &missing spec&comfirmed\\
28 & 6e4d3e389e  & SQLite & Bug &violate spec&comfirmed\\
29 & 411bce39d0  & SQLite & Inconsistency &missing spec&comfirmed\\
\midrule
30 & 6804 & DuckDB & Bug &violate spec&fixed\\
\midrule
31 & 2104 & OceanBase & Inconsistency & missing spec & confirmed\\
32 & 2105 & OceanBase & Inconsistency & missing spec & confirmed\\
\bottomrule 
\end{tabular} 
\end{table}
\subsection{Experiment results}

\noindent\textbf{RQ1: Bugs and inconsistencies. } 
We ran \tool{} on four RDBMSs for a period of 3 months and reported the detected issues to the corresponding developer communities.
Table~\ref{tab:bugs} shows the details of the  confirmed bugs and inconsistencies in four RDBMSs detected by \tool{}. 
We have submitted 32 issues and 19 of them are confirmed by the developers as bugs. 
Out of the issues identified, 23 are related to scalar expressions and 9 to other keywords, including joins and various relational operators. Our primary objective is to detect inconsistencies between RDBMS implementations and the SQL specification by generating test cases that achieve high coverage across different SQL keywords. 
Our implementation focuses on scalar expressions, query expressions, and predicates. Among these, scalar expressions are the most complex, as they often involve combinations of multiple keywords and subqueries. They are also under-specified in the SQL standard and insufficiently tested by existing approaches~\cite{rigger2020finding,rigger2020detecting,rigger2020testing,liang2022detecting}. In contrast, query expressions and predicates are clearly defined in the SQL specification, leading to fewer ambiguities across RDBMSs. Moreover, they have been extensively tested by prior research~\cite{rigger2020finding,rigger2020detecting,rigger2020testing,liang2022detecting}, making it more challenging to uncover new inconsistencies.

Among the confirmed bugs, 1 has been reported previously and 1 has been fixed. The remaining 13 issues are confirmed as inconsistencies, and the developers claim that they were their design choices. 
Among the 13 inconsistencies, 2 of them are due to violation of the SQL specification and 11 of them are due to unclear descriptions in the SQL specification. Developers of different RDBMSs could have different interpretations on the SQL specification, and thus design and implement their RDBMS differently. 
Among the nine inconsistencies arising from unclear standard descriptions, eight inconsistencies were detected in both MySQL (109147, 109148, 109836, 109845, 109962, 110711) and TiDB (42773, 40995). The queries triggering the inconsistencies have  
the same results when executed in MySQL, TiDB, and MariaDB databases, and are different from that of PostgreSQL.
%
It is noteworthy that \textit{all four databases have been extensively tested by existing methods~\cite{rigger2020testing, rigger2020detecting, rigger2020finding}, yet \tool{} is still able to detect bugs that were not detected by those approaches.} 
By a careful inspection on the detected bugs, we find that most of them indeed violate the SQL specification. 
One of the most representative bugs is related to the mishandling of \texttt{NULL} operands in keyword operations. 
According to the SQL specification, 
``\textit{If the value of one or more, $<$string value expression$>$s, $<$datetime value expression$>$s, $<$interval value expression$>$s, and $<$collection value expression$>$s that are simply contained in a $<$numeric value function$>$ is the \texttt{NULL} value, then the result of the $<$numeric value function$>$ is the \texttt{NULL} value}"~\cite{iso-sql-null}.  
However, MySQL violates the specification by returning non-\texttt{NULL} results when operating on \texttt{NULL} values. One bug in MySQL (110438), three bugs in TiDB (42375, 42377, 42378) and one bug in DuckDB 
 (6804) belong to this category. 
Another representative bug is due to incorrect implicit type conversion on string. When a string is converted to a signed integer type, it is mistakenly converted to a float type. There is no specific description in SQL specification on such cases. We will provide detailed analysis in the case study of section~\ref{sec:casestudy} of this type of bugs.  
%
%
Three bugs in MySQL (109149, 109837, 109842) and three bugs in TiDB (39260, 40995, 40996) belong to this category. 

We also identified two bug in TiDB (39258, 39259) which erroneously handles bitwise operations on negative numbers and in operations on string, one bug in MySQL (109146) which is related to improper handling of newline characters by bitwise operators, and three bugs in SQLite  (7e03a4420a, 3f085531bf, 6e4d3e389e) concerning the handling of data anomalies. 
These bugs are specifically related to the improper handling of large numbers or numbers expressed in scientific notations. 
The SQL specification does not provide detailed description on those particular cases. 
%
The remaining two bugs are about column references on TiDB (42376, 42379). When \texttt{RIGHT JOIN} is used together with the \texttt{FIELD} or \texttt{CONCAT_WS} keywords, if the parameters of the function contain references to a certain column, the result set will miss some data records. 
The inconsistencies we identified can be categorized into two types. The first type of inconsistency violates the SQL specification, and the second type of inconsistency arises from variations in the implementations across different RDBMSs due to missing or unclear descriptions in the SQL specification. 
One inconsistency in MySQL (109962) and one in TiDB (42773) are due to the representation of integer \texttt{0} in certain numerical functions, where the result of the integer \texttt{0} is represented as \texttt{-0} because of incorrect implicit type conversion. 
Two inconsistencies in MySQL (110439, 110711) involve anomalies in the results returned by string functions when handling \textbf{NULL} parameters. This inconsistency aligns with some previously identified bugs and contradicts the semantic descriptions of \texttt{NULL} in the SQL specification~\cite{iso-sql-null, iso-sql-isnull}. 
%
We detected three inconsistencies in MySQL (109836, 109845) and TiDB (40995), 
where the precision retained in floating-point results do not align with the specification. Among them, MySQL (109845) and TiDB (40995) do not obey the floating point precision specified by the parameter when using the round function to process integers. In MySQL (109836), performing arithmetic operations on string-type and numeric-type constants with the same numerical value does not yield consistent floating-point precision. Additionally, bitwise operators in both databases do not consistently return signed integer types when performing operations on negative numbers in MySQL (109147) and OceanBase (2104). 
The remaining inconsistencies relate to handling specific data types in MySQL (109148, 110346) or large numbers in SQLite (411bce39d0).  
The SQL specification misses descriptions on those operations, which results in the inconsistencies. 

\begin{tcolorbox}[boxsep=0.01pt]
\textbf{Answer to RQ1:} 
\tool{} detects 19 bugs (18 newly reported) and 13 inconsistencies, which are all confirmed by developers, in four extensively tested RDBMSs. All detected bugs and inconsistencies are either due to RDBMS violating the SQL specification, or missing or unclear SQL specification.  
\end{tcolorbox}

    \begin{figure*}[t]
        \centering
        \begin{minipage}{0.16\linewidth}
            \centering
            \includegraphics[width=\linewidth]{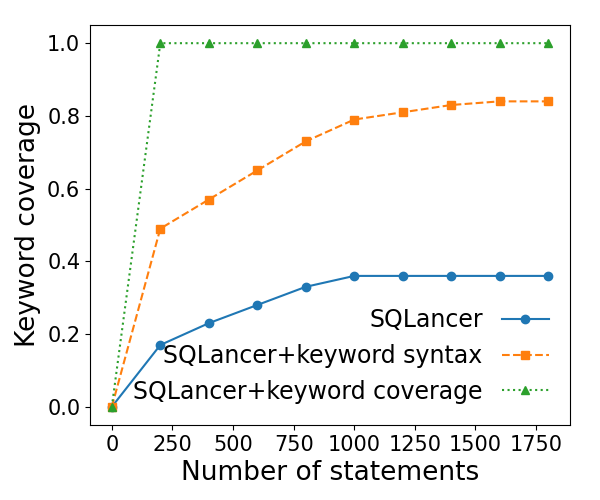}
            \subcaption{MySQL}
            \label{fig:mysql_keyword}
        \end{minipage}
        \begin{minipage}{0.16\linewidth}
            \centering
            \includegraphics[width=\linewidth]{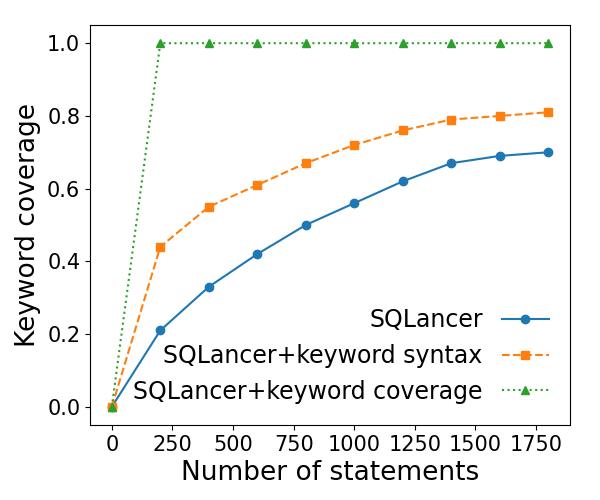}
            \subcaption{TiDB}
            \label{fig:tidb_keyword}
        \end{minipage}
        \begin{minipage}{0.16\linewidth}
            \centering
            \includegraphics[width=\linewidth]{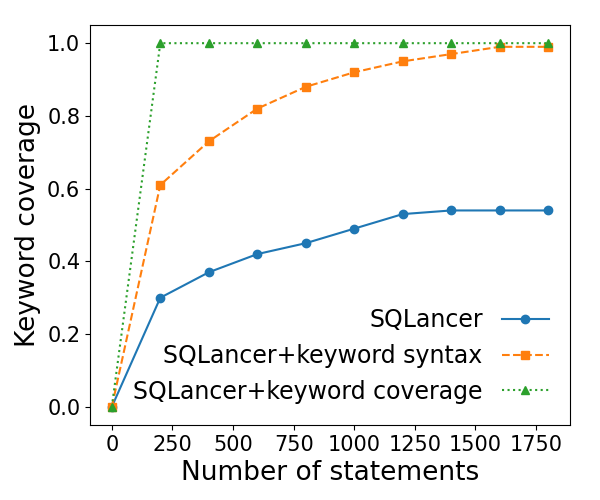}
            \subcaption{SQLite}
            \label{fig:sqlite_keyword}
        \end{minipage}
        \begin{minipage}{0.16\linewidth}
            \centering
            \includegraphics[width=\linewidth]{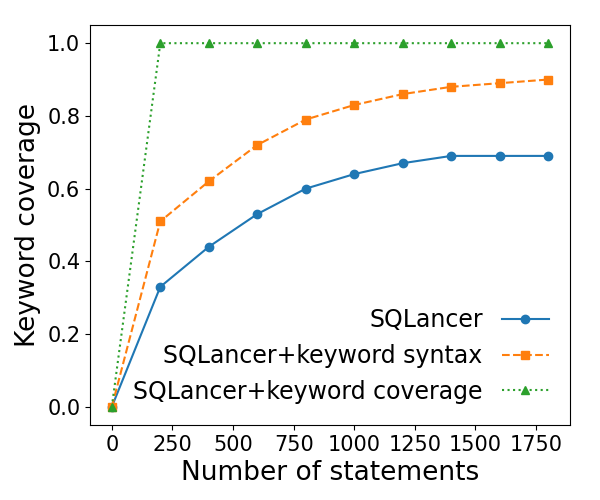}
            \subcaption{DuckDB}
            \label{fig:duckdb_keyword}
        \end{minipage}
        \begin{minipage}{0.16\linewidth}
            \centering
            \includegraphics[width=\linewidth]{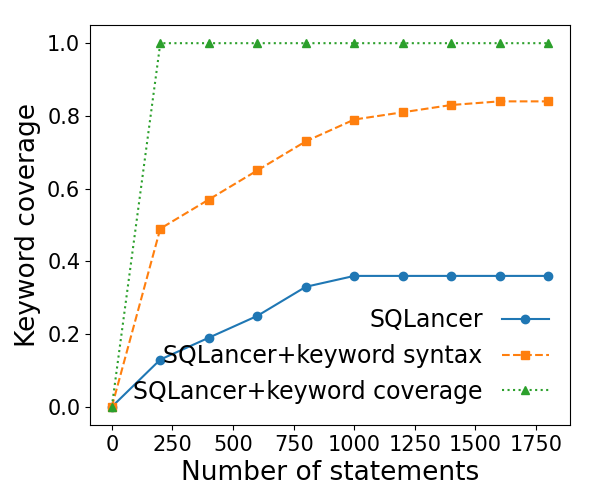}
            \subcaption{OceanBase}
            \label{fig:oceanbase_keyword}
        \end{minipage}
        \begin{minipage}{0.16\linewidth}
            \centering
            \includegraphics[width=\linewidth]{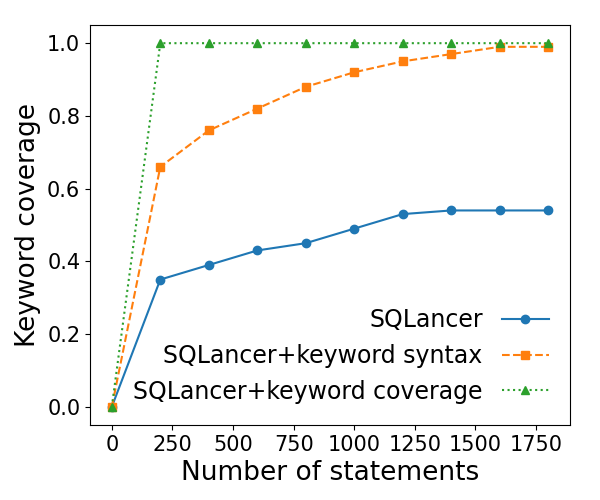}
            \subcaption{PostgreSQL}
            \label{fig:psql_keyword}
        \end{minipage}
        \caption{The keyword coverage increment (y-axis) with the
number of queries (x-axis)} 
\label{fig:keyword_coverage_improvement}
    \end{figure*}

    \vspace{1mm} 

    \begin{figure*}[t]
        \centering
        \begin{minipage}{0.16\linewidth}
            \centering
            \includegraphics[width=\linewidth]{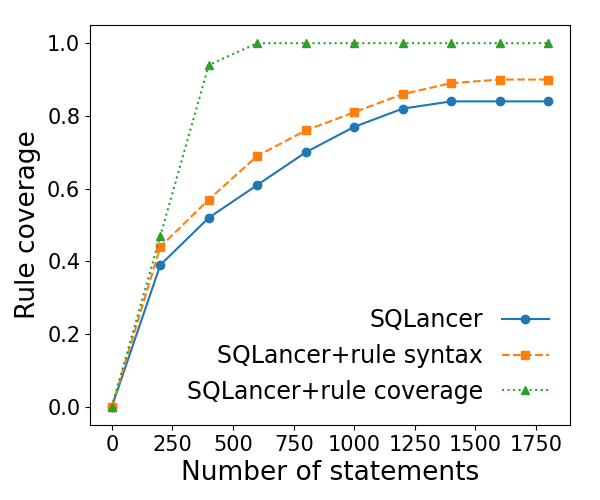}
            \subcaption{MySQL}
            \label{fig:mysql_rule}
        \end{minipage}
        \begin{minipage}{0.16\linewidth}
            \centering
            \includegraphics[width=\linewidth]{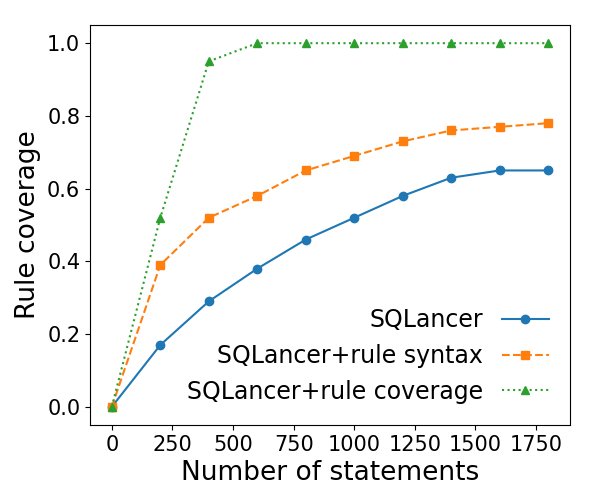}
            \subcaption{TiDB}
            \label{fig:tidb_rule}
        \end{minipage}
        \begin{minipage}{0.16\linewidth}
            \centering
            \includegraphics[width=\linewidth]{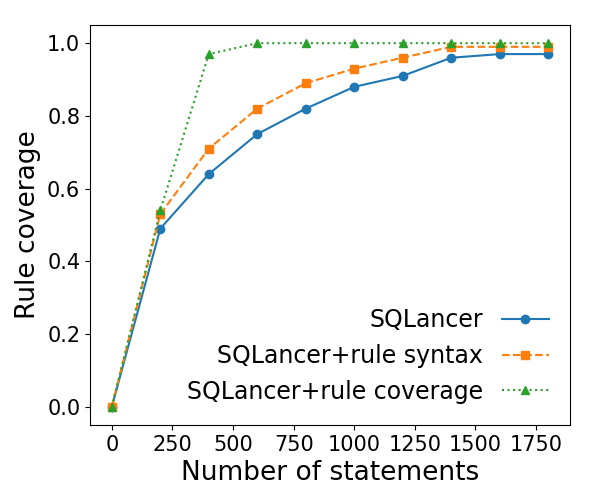}
            \subcaption{SQLite}
            \label{fig:sqlite_rule}
        \end{minipage}
        \begin{minipage}{0.16\linewidth}
            \centering
            \includegraphics[width=\linewidth]{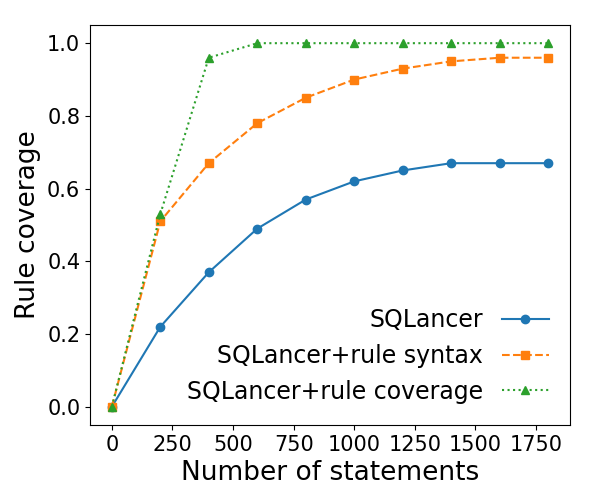}
            \subcaption{DuckDB}
            \label{fig:duckdb_rule}
        \end{minipage}
        \begin{minipage}{0.16\linewidth}
            \centering
            \includegraphics[width=\linewidth]{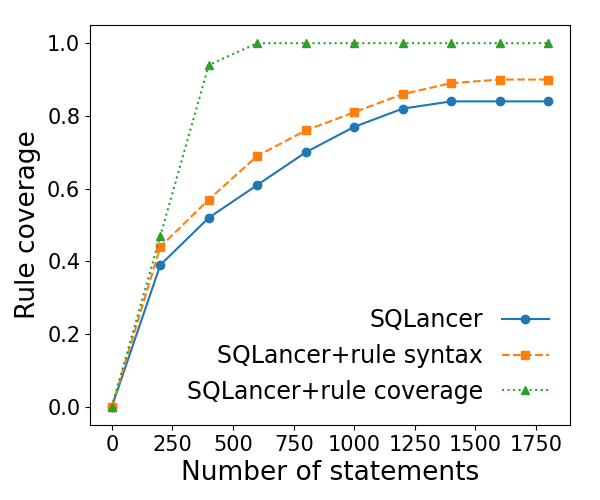}
            \subcaption{OceanBase}
            \label{fig:oceanbase_rule}
        \end{minipage}
        \begin{minipage}{0.16\linewidth}
            \centering
            \includegraphics[width=\linewidth]{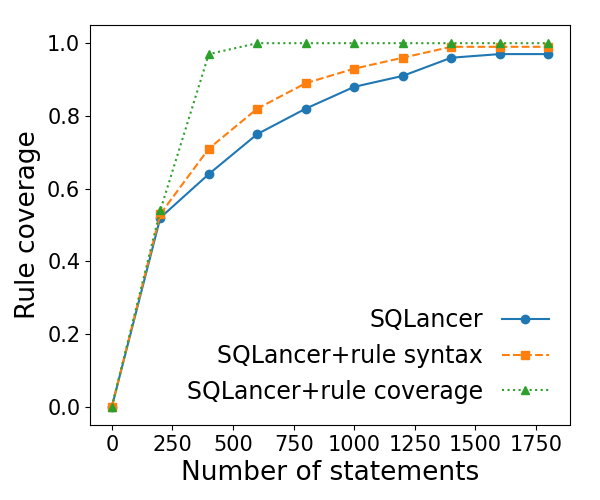}
            \subcaption{PostgreSQL}
            \label{fig:psql_rule}
        \end{minipage}
        \caption{The rule coverage increment (y-axis) with the number of queries (x-axis)}
        \label{fig:Rule_coverage_improvement}
    \end{figure*}

    \vspace{1mm} 

    \begin{figure*}[t]
        \centering
        \begin{minipage}{0.16\linewidth}
            \centering
            \includegraphics[width=\linewidth]{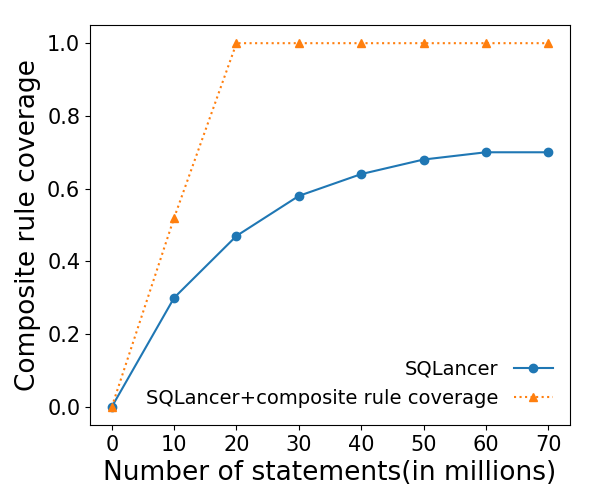}
            \subcaption{mysql}
            \label{fig:mysql_cr}
        \end{minipage}
        \begin{minipage}{0.16\linewidth}
            \centering
            \includegraphics[width=\linewidth]{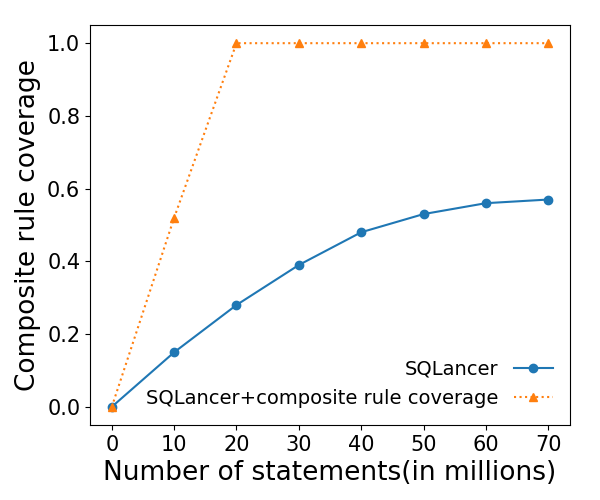}
            \subcaption{TiDB}
            \label{fig:tidb_cr}
        \end{minipage}
        \begin{minipage}{0.16\linewidth}
            \centering
            \includegraphics[width=\linewidth]{img/Coverage_Impovement/sqlite_rule.png}
            \subcaption{SQLite}
            \label{fig:sqlite_cr}
        \end{minipage}
        \begin{minipage}{0.16\linewidth}
            \centering
            \includegraphics[width=\linewidth]{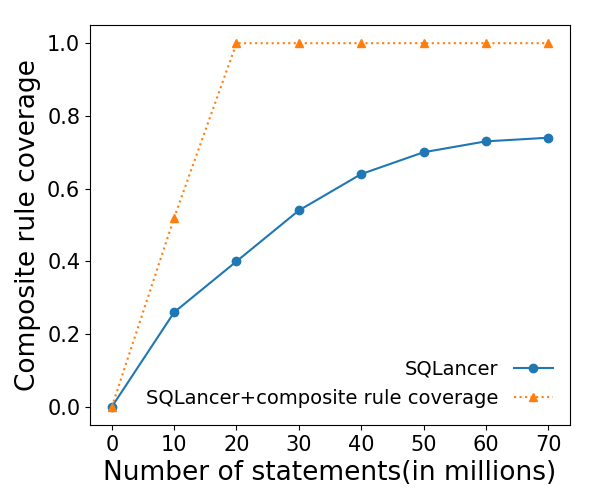}
            \subcaption{DuckDB}
            \label{fig:duckdb_cr}
        \end{minipage}
        \begin{minipage}{0.16\linewidth}
            \centering
            \includegraphics[width=\linewidth]{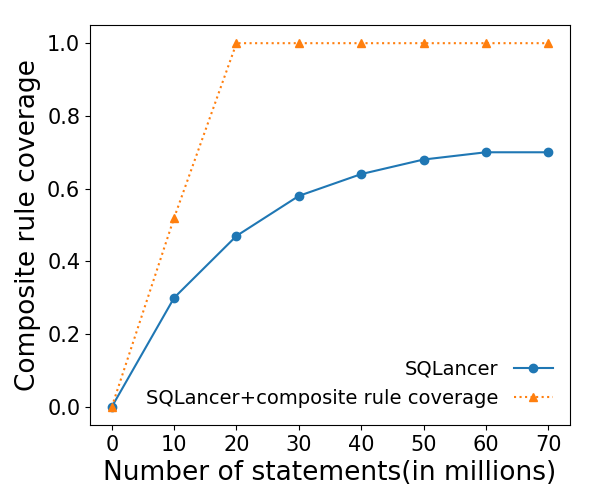}
            \subcaption{OceanBase}
            \label{fig:oceanbase_cr}
        \end{minipage}
        \begin{minipage}{0.16\linewidth}
            \centering
            \includegraphics[width=\linewidth]{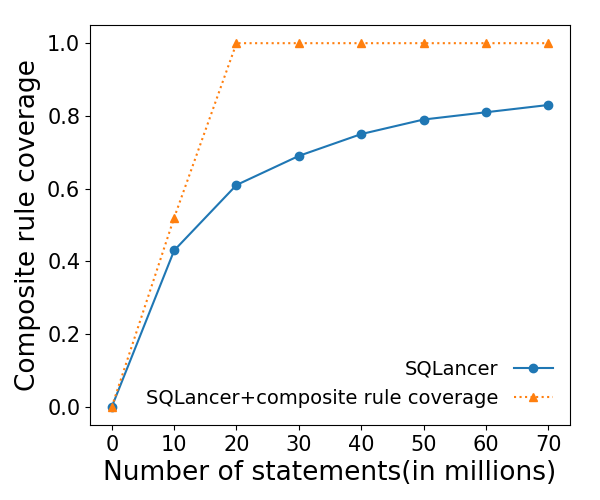}
            \subcaption{PostgreSQL}
            \label{fig:psql_cr}
        \end{minipage}
    
    \caption{The composite rule coverage increment (y-axis)
with the number of queries (in million).} 
    \label{fig:Composite_rule_improvement}
\end{figure*}



\noindent\textbf{RQ2: Effectiveness of the coverage criteria}?
We measure the effectiveness of the proposed coverage criteria in two aspects, i.e., whether they are effective in guiding generating test cases that achieve higher coverage, and whether they are effective in guiding generating test cases that uncover unknown bugs or inconsistencies. 
The experiment results on keyword coverage, rule coverage and composite rule coverage  improvement are shown in Figure~\ref{fig:keyword_coverage_improvement}, Figure~\ref{fig:Rule_coverage_improvement}  and 
Figure~\ref{fig:Composite_rule_improvement}, respectively. 

In Figure~\ref{fig:keyword_coverage_improvement}, 
SQLancer+keyword syntax represents the setting of adding keywords and the corresponding generation rules which were not supported by SQLancer. 
SQLancer+keyword syntax greatly improved the keyword coverage for all four databases. 
Notably, within the first 1500 SQL statements, over 80\% of the keywords were covered on all four databases, with SQLite achieving an impressive keyword coverage of 99\%. 
Keyword coverage guided query generation (SQLancer+keyword coverage) further improves the keyword coverage, and achieved 100\% keyword coverage within the first 200 generate queries for all databases, demonstrating the effectiveness of our keyword-guided query generation method. 

\begin{table*}[h]
\footnotesize
\setlength{\tabcolsep}{5pt}
\centering
\caption{The bug and inconsistency numbers detected by \tool{} with no coverage guided and coverage guided in 6h}
\label{table:The_bug_numbers_detected_by_our_tool_with/without_coverage_in_6h}
\begin{tabular}{ccccccccccccc}
\toprule
\multirow{2}{*}{DBMS} & \multicolumn{3}{c}{\tool{}} & \multicolumn{3}{c}{\tool{}+keyword coverage} & \multicolumn{3}{c}{\tool{}+rule coverage} & \multicolumn{3}{c}{\tool{}+composite rule coverage} \\
\cmidrule{2-13}
& Bugs & Inconsistencies & Time & Bugs & Inconsistencies & Time & Bugs & Inconsistencies & Time & Bugs & Inconsistencies & Time \\
\midrule
MySQL & 3 & 3  & 12.42 & 4 & 5 & 11.25 & 5 & 6 & 5.75& 5 & 6 & 6.23   \\
TiDB & 3 & 1 & 13.33 & 6 & 2 & 12.17& 5 & 2 & 7.46 & 7 & 2 & 8.62  \\
SQLite & 1 & 1 & 20.50 & 2 & 1 & 17.83& 3 & 1 & 14.33 & 3 & 1 & 13.05  \\
DuckDB & 1 & 0 & 29.37 & 1 & 0 & 27.32 & 1 & 0& 31.68& 1 & 0 & 26.29  \\
OceanBase & 0 & 2 & 13.25 & 0 & 2 & 13.37 & 0 & 2& 7.92& 0 & 2 & 9.13  \\
\bottomrule
\end{tabular}
\end{table*}



\begin{table*}[t]
\footnotesize
\centering
\caption{The bugs and inconsistencies detected by SQLancer, SQLRight and \tool{} in 6h}
\label{table:The_bug_numbers_detected_by_SQLancer_and_our_tool_in_72h}
\begin{tabular}{ccccccccccc}
\toprule
\multirow{2}{*}{DBMS} & \multicolumn{2}{c}{TLP (SQLancer)} & \multicolumn{2}{c}{NoREC (SQLancer)} & \multicolumn{2}{c}{TLP (SQLRight)} & \multicolumn{2}{c}{NoREC (SQLRight)} & \multicolumn{2}{c}{\tool{}} \\
\cmidrule{2-11}
& Bugs & Inconsistencies & Bugs & Inconsistencies & Bugs & Inconsistencies & Bugs & Inconsistencies & Bugs & Inconsistencies \\
\midrule
MySQL & 1 & 0 & - & - & 1 & 0 & 0 & 0 & 5 & 6 \\
TiDB & 2 & 0 & - & - & - & - & - & - & 7 & 2 \\
SQLite & 0 & 1 & 0 & 0 & 0 & 0 & 0 & 0 & 3 & 1 \\
DuckDB & 1 & 0 & 0 & 0 & - & - & - & - & 1 & 0 \\
OceanBase & 0 & 1 & 0 & 0 & - & - & - & - & 0 & 2 \\
\bottomrule
\end{tabular}
\vspace{-5mm}
\end{table*}

Figure~\ref{fig:Rule_coverage_improvement} shows the results on rule coverage, which show similar trend with that on keyword coverage. 
Due to the limited support of SQL features, e.g., data types, by SQLancer, especially for DBMS such as DuckDB and TiDB, relying only on SQLancer achieves low rule coverage, as shown in Figure~\ref{fig:Rule_coverage_improvement}. Therefore, we add those missing features in SQLancer for the corresponding DBMS query generation and refer this as SQLancer + rule syntax.
We can observe that adding those missing features improves the rule coverage, especially for DuckDB and TiDB. Rule coverage-guided query generation (SQLancer+rule coverage)  achieves the highest rule coverage with the fewest number of queries. The results indicate the effectiveness of our rule coverage-guided query generation algorithm. 
Figure~\ref{fig:Composite_rule_improvement} depicts the improvements in composite rule coverage by the generated queries for the four databases. 
With SQLancer, which conducts random query generation, we observed that the increase in composite rule coverage tends to plateau after generating 60 million data points. 
At this stage, MySQL, SQLite, DuckDB and OceanBase each achieved a composite rule coverage of around 70\% and TiDB 50\%. 
With the introduction of composite rule coverage guidance, all four databases were able to reach 100\% composite rule coverage after generating 20 million queries (our Prolog implementation has a total of 19 million composite rules). The results indicate the effectiveness of our composite rule coverage-guided query generation algorithm. 


To verify the effectiveness of the coverage-guided query generation algorithm in assisting detecting bugs and inconsistencies in relational DBMS, we conducted an ablation study of \tool{} with and without coverage guidance. 
Table~\ref{table:The_bug_numbers_detected_by_our_tool_with/without_coverage_in_6h}] presents the experimental results obtained from testing four databases over a period of 6 hours.
%
%
We record the number of bugs and inconsistencies detected on the four settings, i.e.,  \tool{} without coverage guidance, and \tool{} with three coverage guidance. We also report the time taken to discover the first bug or inconsistency. 
Note that to conduct fair comparisons, we improved SQLancer by incorporating all keywords supported by our semantics and related generation rules in \tool{}.
The experimental results indicate that within a 6-hour timeframe, all three coverage metrics successfully assist detecting more bugs and inconsistencies compared with random generation. Composite rule coverage is the most effective among all three coverage metrics. 
In terms of the time taken to detect the first bug or inconsistency, all three coverage guidance algorithm are faster than \tool{} with random query generation. 
In particular, keyword coverage, rule coverage and composite rule coverage are 5.22\%, 24.19\% and 21.41\% faster than random query generation.

\begin{tcolorbox}[boxsep=0.01pt]
\textbf{Answer to RQ2:} 
The three coverage criteria all improve the query generating process, triggering more bugs and inconsistencies  with faster speed. Composite rule coverage achieves  the most significant improvement. 
\end{tcolorbox}

\noindent\textbf{RQ3: How does \tool{} perform compared to baseline approaches}?

We compare \tool{} with two state-of-the-art approaches TLP~\cite{rigger2020detecting} and NoREC~\cite{rigger2020finding}, which are metamorphic testing approaches for relational DBMS. For both approaches, we adopt SQLancer~\cite{sqlancer} and SQLRight~\cite{liang2022detecting} for query generation.  
Notably, SQLancer does not support the NoREC oracle  for MySQL and TiDB, while SQLRight does not support TiDB, DuckDB and OceanBase. 
Therefore, we excluded these specific scenarios from our experiments. 
We ran the compared tools for a period of 6 hours and  
report the results in Table~\ref{table:The_bug_numbers_detected_by_SQLancer_and_our_tool_in_72h}.

The experimental results show that both SQLancer and SQLRight using the NoREC as the oracle were unable to detect new bugs or inconsistencies. 
The TLP oracle with SQLancer for query generation detected 4 bugs in three databases, and with SQLRight for query generation detectd 1 bug in MySQL.
\tool{} outperformed the compared approaches and detected 16 bugs and 9 inconsistencies in  the four databases. 
The reason is that existing approaches do not consider the SQL specification and thus fail to find bugs that violated the SQL specification. 
For instance, One bug (109842) we detected in MySQL is related to the \texttt{MOD} function. 
When applied to negative numbers, MySQL incorrectly represents the result as \texttt{-0}. 
Both TLP and NoREC failed to detect this bug, even after successfully generated the bug triggering query. 

On average, our tool finds a bug in 67 minutes using 19,381 test cases, compared to 300 minutes and 1.3 million test cases for SQLancer, and 30 hours and 8.4 million test cases for SQLRight.  
Our approach finds more bugs/inconsistencies with fewer test cases, demonstrating its effectiveness and efficiency in detecting bugs and inconsistencies that violating SQL specifications.

\begin{figure}[h]
    \begin{minipage}{0.45\linewidth}
        \centering
        \includegraphics[width=\linewidth]{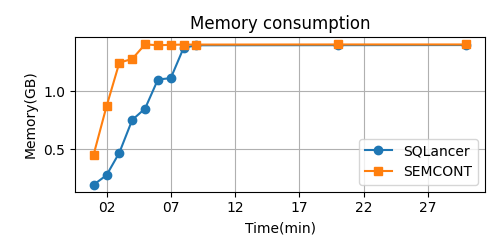}
        \subcaption{MySQL}
        \label{fig:MySQL_memory}
    \end{minipage}%
    \begin{minipage}{0.45\linewidth}
        \centering
        \includegraphics[width=\linewidth]{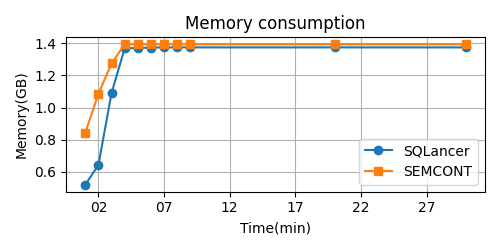}
        \subcaption{TiDB}
        \label{fig:TiDB_memory}
    \end{minipage}

    \begin{minipage}{0.45\linewidth}
        \centering
        \includegraphics[width=\linewidth]{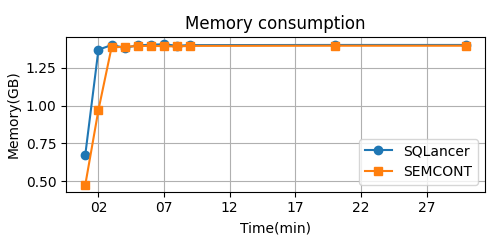}
        \subcaption{SQLite}
        \label{fig:SQLite_memory}
    \end{minipage}%
    \begin{minipage}{0.45\linewidth}
        \centering
        \includegraphics[width=\linewidth]{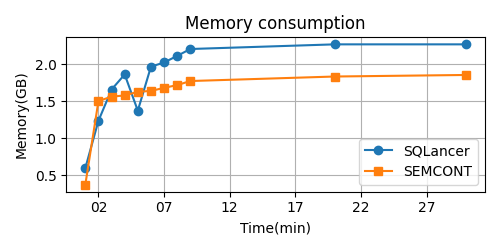}
        \subcaption{DuckDB}
        \label{fig:DuckDB_memory}
    \end{minipage}

    \begin{minipage}{0.45\linewidth}
        \centering
        \includegraphics[width=\linewidth]{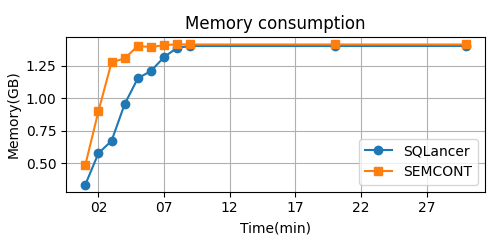}
        \subcaption{OceanBase}
        \label{fig:oceanbase_memory}
    \end{minipage}
    \begin{minipage}{0.45\linewidth}
        \centering
        \includegraphics[width=\linewidth]{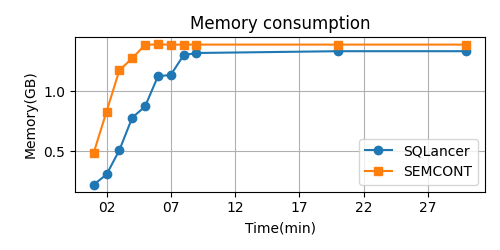}
        \subcaption{PostgreSQL}
        \label{fig:psql_memory}
    \end{minipage}
    \caption{The memory consumption of SEMCONT with SQLancer.}
    \label{fig:memory1}
\end{figure}

\begin{figure}[t]
\footnotesize
\centerline{\includegraphics[width=0.5\textwidth,height=0.16\textwidth]{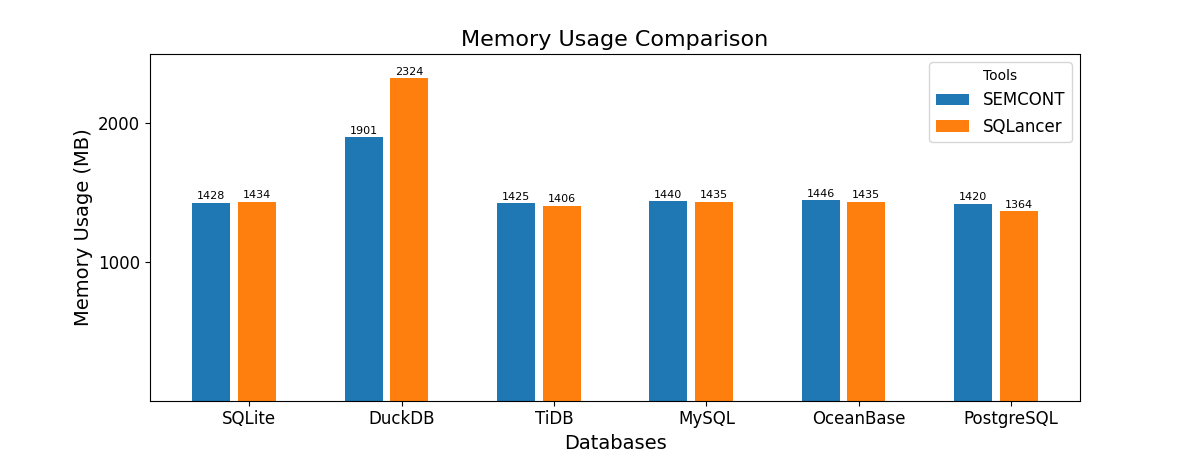}}
\caption{Memory consumption of \tool{} and SQLancer} 
\label{fig:memory}
\vspace{-7mm} 
\end{figure}

\noindent\textbf{Memory Consumption.}
Figure~\ref{fig:memory1} and Figure~\ref{fig:memory} shows the memory usage of \tool{} and SQLancer during a 6-hour test on six databases. Memory usage is mainly influenced by query generation, execution, and result comparison, with query generation being the most memory-intensive. We report stabilized memory consumption, with detailed time-based changes provided in our technical report~\cite{technicalReport}. Results indicate that \tool{}’s memory usage is comparable to the baseline. Memory consumption is low because we use small tables (a few hundred records). We do not test concurrent query execution, resulting in stable memory usage.

Our goal is to detect compliance bugs in RDBMSs, not to evaluate performance. \tool{} is designed for offline testing during RDBMS development, and its overhead is manageable.
\tool{} shows very low memory consumption and it requires far less test cases to detect a compliance issue, demonstrating its efficiency.

\begin{tcolorbox}[boxsep=0.01pt]
\textbf{Answer to RQ3:} 
\tool{} detected more bugs and inconsistencies than the compared baseline approaches. Baseline approaches fail to detect those bugs since they do not refer to SQL specification in their testing process. 
\end{tcolorbox}

\noindent\textbf{RQ4: Case study. }
\label{sec:casestudy}
%
%
\tool{} has identified 19 bugs and 13 inconsistencies, which  
arise from two reasons, i.e., (1) DBMS implementations are not consistent with SQL specification and (2) unclear or missing description in the SQL specification.  
These problems have resulted in variations in the specific implementations across different RDBMSs, consequently affecting the user experience. 

\begin{figure}[t]
\centering
\begin{lstlisting}
@\normalcolor{SELECT}@ MOD('-12',-4); 
--expected: 0 @\color{green}{\CheckmarkBold}@, actual:-0 @\color{red}{\XSolidBrush}@
\end{lstlisting}
\caption{A bug in MySQL 8.0.29 }
\label{fig:bug2}
\end{figure}

\noindent\textbf{A bug due to missing specifications.} 
Figure~\ref{fig:bug2} is a bug we detected in MySQL 8.0.29. The query conducts  \texttt{MOD} function with the string type as the first parameter. The expected result of the query is \texttt{0}, yet MySQL returned \texttt{-0}. 
MySQL developers confirmed this bug and explained the reason is that when the first parameter of \texttt{MOD} is a string type, an implicit type conversion should be triggered to convert the string type \texttt{'-12'} to a signed integer \texttt{-12}. However, MySQL mistakenly converted \texttt{'-12'} to a float type \texttt{-12.0}, causing this bug.  
SQL specification does not specify how to convert a string type to a numeric type, and thus different RDBMSs may make on their own implementation choices. 

\vspace{1mm}
\noindent\textbf{An inconsistency violating the SQL specification.}
Figure~\ref{fig:NULL} shows the queries that cause an inconsistency in MySQL 8.0.29 that violates the SQL specification. 
The first three SQL queries create three tables \texttt{t0}, \texttt{t1} and \texttt{t2}. 
Then \texttt{t0} and \texttt{t1} are inserted values \texttt{NULL} and string \texttt{`hhhh'} (lines 4, 5), respectively. Line 6 replaces the value in \texttt{t2} with value \texttt{960364164}. 
The query in line 7 returns an empty list in MySQL 8.0.29, which violates the SQL specification~\cite{iso-sql-isnull}.





\begin{figure}[t]
\begin{lstlisting}[language=SQL]
CREATE TABLE IF NOT EXISTS t0(c0 LONGTEXT STORAGE DISK COMMENT 'asdf' COLUMN_FORMAT FIXED) ;
CREATE TABLE IF NOT EXISTS t1 LIKE t0;
CREATE TABLE IF NOT EXISTS t2(c0 DECIMAL ZEROFILL COMMENT 'asdf' COLUMN_FORMAT FIXED PRIMARY KEY UNIQUE STORAGE DISK);
INSERT INTO t0(c0) VALUES(NULL);
INSERT INTO t1(c0) VALUES('hhhh');
REPLACE INTO t2(c0) VALUES(960364164);
SELECT t1.c0, t2.c0 FROM t1, t2 RIGHT OUTER JOIN t0 ON 0 WHERE (NOT ((t2.c0 IS FALSE)!= ((t1.c0))));
-- expected:[[`hhhh',NULL,NULL]], actual:[]
\end{lstlisting}
\caption{The queries triggering an inconsistency in MySQL 8.0.29 with the SQL specification}
\label{fig:NULL}
\end{figure}

The select query in line 7 involves a right outer join between \texttt{t2} and \texttt{t0} on condition \texttt{0}, meaning \texttt{false} in this context, and thus no matching columns are returned from the two tables. Therefore, the resulting table will retain all the data from the right table (t0) and replace all data from the left table (t2) with \texttt{NULL} for the \texttt{RIGHT OUTER JOIN} operation, and a table with one data record \texttt{[NULL, NULL]} (on columns t0.c0 and t2.c0) is returned. 
Then natural join of table \texttt{t1} with that result table is performed, resulting a table with one record \texttt{['hhhh', NULL, NULL]}. 
The \texttt{WHERE} condition is the tricky part which causes the inconsistency. Since \texttt{t2.c0} is \texttt{NULL} (after the right outer join), the result of \texttt{(t2.c0 IS FALSE)} should be \texttt{FALSE}. According to the SQL specification~\cite{iso-sql-isnull} (the truth table for \texttt{IS BOOLEAN} operator in \textit{Part 2 Foundation, boolean value expression}), the truth value for \texttt{NULL IS FALSE} and \texttt{NULL IS TRUE} should both be false.
On the left of the comparison operator \texttt{!=} is a boolean type and on the right a string type. Therefore, MySQL will convert the boolean type \texttt{false} to a numeric number \texttt{0} and try to convert the string type to an integer type by default. In this case, the first character of the string \texttt{'hhhh'} is a non-numeric character,  it is converted to integer \texttt{0}. 
Therefore, \texttt{(t2.c0 is FALSE) != (t1.c0)} is evaluated to false and thus the \texttt{WHERE} condition is evaluated to true. 
The returned result should be \texttt{['hhhh', NULL, NULL]} according to the SQL specification. Yet MySQL 8.0.29 returned an empty list. We have reported this inconsistency to the MySQL developers and they confirmed the reason for this inconsistency is due to the violation of SQL specification on the truth value of the \texttt{IS FALSE} operator.  
Four inconsistencies that we detected are due to the same reason.

\begin{figure}[t]
\centering
\begin{subfigure}{0.45\textwidth}
\begin{lstlisting} 
SELECT "1gbnn" IN (1); 
   -- PostgreSQL:invalid input syntax for type integer 
   -- TiDB: 1
\end{lstlisting}
\caption{The query that triggers different results in PostgreSQL and TiDB}
\label{fig:inconsistency2}
\end{subfigure}
\begin{subfigure}{0.45\textwidth}
\centering
\includegraphics[width=\textwidth,height=0.15\textwidth]{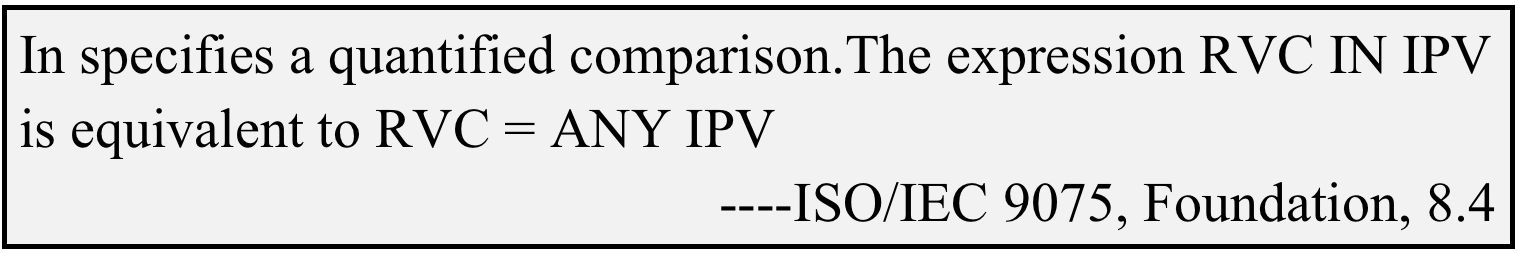}
\caption{The SQL specification about the \texttt{IN} operator}
\label{fig:specification2}
\end{subfigure}
\caption{An inconsistency caused by unclear description of the \texttt{IN} operator}
\label{fig:unclearspec}
\end{figure}

\noindent\textbf{An inconsistency due to unclear SQL specification.}
%
Figure~\ref{fig:inconsistency2} shows a query which triggers different results in PostgreSQL and TiDB. 
PostgreSQL reports an invalid input syntax for type integer, while TiDB returns the result of \texttt{1} (representing \texttt{true} in this context). 
The documentation of TiDB states that data comparison involves implicit type conversion before the actual comparison. 
According to the SQL specification shown in Figure~\ref{fig:specification2}, the \texttt{IN} operator is described as a quantitative comparison operator, but the specification does not clearly instruct on how to conduct comparison on different data types.
Therefore, different RDBMS developers have different implementation choices, resulting in this inconsistency among different RDBMSs, which may confuse the users. 
%
Seven inconsistencies we detected are due to similar reasons. 

We argue that both types of inconsistencies could lead to user confusion, unexpected behavior or even serious consequences in commercial or safety critical scenarios. 
It is thus crucial for RDBMS developers to conform to the SQL specification, and the 
SQL specification should also be improved to avoid potential confusion and misunderstandings.

\begin{table*}[t]
    \centering
    \caption{Inconsistencies between databases}
    \resizebox{\textwidth}{!}{
    \begin{tabular}{ccccccc}
        \toprule
      Inconsistency Description  & Example & MySQL & TiDB & DuckDB & SQLite & PostgreSQL \\
        \midrule
        \multirow{2}{*}{Implicit type conversion problem} & SELECT '1' IN (1) & 1 & 1 & 0 & 0 & 0 \\
         & SELECT 1 \% '1E1'; & 1 & 1 & 1 & 0 & 1 \\
         \hline
        \multirow{3}{*}{The result representation of the bitwise operator}& SELECT \textbackslash n1' \& 1 & 0 & 0 & 1 & 1 & 1 \\
        & SELECT -5 \& -4; & 18446744073709551608 & 18446744073709551608 & -8 & -8 & -8 \\
        & SELECT -3 $>$ ('5' $^{}$ '-4'); & 0 & 0 & 1 & 1 & 1 \\
        \hline
        \multirow{2}{*}{The result representation of zero}& SELECT '0'/-4; & -0 & -0 & 0 & 0 & 0 \\
        & SELECT mod('-12',-4); & -0 & -0 & 0 & 0 & 0 \\
        \hline
        The result representation of the round function& SELECT round(1,2); & 1 & 1 & 1.00 & 1.00 & 1.00 \\
        \hline
        The result representation of floating point& SELECT '1'/32; & 0.03125 & 0.03125 & 0.313 & 0.313 & 0.313 \\
        \bottomrule
    \end{tabular}
    }
    \label{Inconsistencies between databases}
\end{table*}

Table~\ref{Inconsistencies between databases} shows the inconsistencies we found between databases due to unclear descriptions of the standards.

\begin{tcolorbox}[boxsep=0.01pt]
\textbf{Answer to RQ4:} 
Among the 19 bugs and 13 inconsistencies we detected, 8 bugs and 2 inconsistencies are due to violating the SQL specification, 11 bugs and 11 inconsistencies are due to unclear or missing descriptions in the SQL specification. Both types of inconsistencies could confuse users and should be carefully handled. 
 
\end{tcolorbox}

\subsection{Threats to Validity}
\label{sec:threatstovalidity}

\noindent\textbf{Coverage of the semantics. }
We defined the denotational semantics of all keywords concerning the SQL Data Query Language (DQL) commands. For the other types of SQL commands, including DDL, SML, DCL and TCL, they can be easily supported by extending our semantics. 

\noindent\textbf{Randomness. }
The number of bugs/inconsistencies detected as well as the coverage increase could be affected by the randomness of test query generation. To address the issue, we run all comparison  experiments, i.e., RQ2 and RQ3, 5 times and report the median number. Moreover, in the comparison experiments of different settings of \tool{} (Table~\ref{table:The_bug_numbers_detected_by_our_tool_with/without_coverage_in_6h}) and comparison with baselines (Table~\ref{table:The_bug_numbers_detected_by_SQLancer_and_our_tool_in_72h}), we run each experiment for 6 hours to mitigate the effect of randomness. 

\noindent\textbf{Correctness of \tool{}'s implementation. }
Firstly, we conduct unit test on every semantic rule we implemented. We also conduct integrated testing by generating complex queries to test the composite semantic rules. Moreover, two of the co-authors conduct manual code inspection independently to ensure the correctness of \tool{}'s implementation. 
Another issue is when the SQL specification does not clearly describe the semantics of a keyword, operator or function. In this scenario, we need to make our own design choices. We first refer to SQL specification for similar operators, which have clear semantic descriptions. 
If such operators are not found, we check the implementation of existing RDBMSs and adopt the one that majority RDBMSs use. 
We have encounter 47 such cases.

\vspace{1mm}
\noindent\textbf{Fairness of experiment design.} We select two state-of-the-art metamorphic testing approaches for RDBMS as baselines. Moreover, for each baseline, we use two state-of-the-art query generation methods, which result in four settings of the compared approach. We follow the experiment settings of the baselines presented in their papers to reproduce their best performance. All experiments are conducted on the same environment, within the same time frame and the same configurations of the tested RDBMSs for fairness of comparison. 
\section{Related Work}
\label{sec:relatedwork}

\noindent\textbf{Testing Relational DBMS}
\label{sec:testing approach}
Research on testing relational database systems to identify and rectify logical bugs is a pivotal and actively pursued area of study in database management systems. 
Existing methods for testing relational database systems primarily encompass two distinct categories, metamorphic testing, which focuses on validating the database's behavior under varying conditions without known output~\cite{journals/corr/abs-2002-12543,journals/tse/ZhouXC16,journals/tse/SeguraFSC16,journals/csur/ChenKLPTTZ18}, and differential testing, where the system's responses to identical inputs are compared across different versions or configurations.

RAGS~\cite{slutz1998massive} and Apollo~\cite{jung2019apollo} are notable early method that implemented differential testing to identify logical errors in database management systems.
SQLSmith~\cite{seltenreich2019sqlsmith} employs a technique of continuously generating random SQL query statements for database testing. 
However, this method is limited to identifying defects that lead to system crashes, and does not cover other types of logical errors.
Ratel~\cite{wang2021industry} significantly improves the robustness of SQL generation for database testing by merging SQL dictionaries with grammar-based mutations. 
Furthermore, it boosts the accuracy of feedback through the use of binary coverage linking and bijective block mapping, thereby enhancing the overall performance of RDBMS testing.
SparkFuzz~\cite{ghit2020sparkfuzz} introduces a fuzzing-based method that utilizes the query results from a reference database as test oracles.
The effectiveness of these differential testing methods is limited by the shared functionalities and syntax supported across the databases under test. 
Moreover, they may yield false positives due to the varied implementation choices inherent in different database systems.

Numerous testing methods involve constructing a test oracle by proposing a variety of metamorphic relations~\cite{ghit2020sparkfuzz,chen2020testing,rigger2020testing,rigger2020detecting,rigger2020finding}.
MUTASQL~\cite{chen2020testing} and Eqsql~\cite{castelein2018search} are tools that construct test cases by defining mutation rules. 
These rules are used to generate or synthesize SQL query statements that are functionally equivalent to the original ones.
In recent years, SQLancer~\cite{sqlancer} has emerged as the most effective black-box fuzz testing tool, distinguished by its adoption of three complementary oracles~\cite{rigger2020testing,rigger2020detecting,rigger2020finding}.
PQS~\cite{rigger2020testing} operates by first selecting a row of data, and then synthesizing a query based on this selected data. 
The design of the query is such that it must return the initially chosen row. 
This approach detects bugs by verifying whether the returned result includes the specific row of data.
NoREC~\cite{rigger2020detecting} transforms an optimized SQL query into an equivalent non-optimized version and then compares the execution results of both.
TLP~\cite{rigger2020finding} divides a SQL query into three separate SQL statements that collectively retain the same semantics as the original query. 
If the results of executing the original query differ from those obtained from the three divided queries, it likely indicates the presence of logical errors.
SQLRight~\cite{liang2022detecting} focuses on enhancing the semantic correctness of generated SQL queries and adopts the oracles proposed by PQS~\cite{rigger2020testing}, NoREC~\cite{rigger2020detecting}, and TLP~\cite{rigger2020finding}.
GRIFFIN~\cite{fu2022griffin} executes mutation testing within the grammatical boundaries of SQL language, transforming data into metadata for this purpose.
While metamorphic testing approaches address syntax differences arising from various database implementations, consistently returned results do not always guarantee the absence of bugs. 
Furthermore, these methods fall short in detecting bugs caused by violations of SQL specifications.

\vspace{1mm}
\noindent\textbf{Formal semantics for SQL. }
%
There have been several approaches ~\cite{ceri1985translating, van2009translating, negri1991formal, chu2017cosette, chu2017hottsql, malecha2010toward, veanes2010qex} that made attempts to formalize the semantics of SQL. 
Chinaei~\cite{chinaei2007ordered} was the first to propose a bag-based SQL operational semantics. 
More recent works~\cite{ricciotti2022formalization, guagliardo2017formal, benzaken2019coq} have taken into account NULL values when defining SQL formal semantics. 
Guagliardo and Libkin~\cite{guagliardo2017formal} defined the formal semantics of SQL, taking into account not only the syntax of basic SQL query statements but also data structures such as subqueries, sets, and bags, which were not supported in previous research. 
This work also implemented and verified the correctness of its operational semantics in programming. 
SQLcoq~\cite{benzaken2019coq} delves into grouping and aggregate functions, proving the equivalence between its proposed formal semantics and relational algebra. 
Additionally, Zhou et al.~\cite{zhou2020symbolic} introduces an algorithm for proving query equivalence under bag semantics. 
These methods~\cite{guagliardo2017formal, benzaken2019coq, zhou2020symbolic, cheney2021comprehending} have significantly enhanced the formal definition of SQL by comprehensively considering both bag and NULL semantics.
However, 
these methods support only a subset of the functionalities defined in the SQL specification.
Moreover, these work did not apply semantics for database conformance testing, since these semantics are primarily developed for correctness verification rather than efficient automatic testing.

\vspace{1mm}
\noindent\textbf{Semantics based testing. }
Efforts have also been made to utilize executable semantics as test oracles~\cite{roșu2010overview, schumi2022exais}. 
Various popular programming languages, developed using the K framework~\cite{roșu2010overview}, offer executable semantics that can be effectively used as testing oracles. 
ExAIS~\cite{schumi2022exais} formalizes executable semantics for artificial intelligence libraries and implements them using the Prolog language. 
These semantics are not only employed as testing oracles but also play a pivotal role in guiding test case generation. 
Our work presents the first initiative to use semantics for testing in the domain of database management systems and detect 19 bugs and 11 inconsistencies, most of which cannot be detected by existing approaches. 

\section{CONCLUSION}

In this work, we propose the first automatic conformance testing approach, \tool{}, for RDBMSs with the SQL specification. We define the formal semantics of SQL and implement them in Prolog, which then act as the oracle for the conformance testing. Moreover, we define three coverage criteria based on the formal semantics to guide test query generation. The evaluation with four well known and extensively tested RDBMSs show that, \tool{} detects 19 bugs (18 of which are reported for the first time) and 13 inconsistencies, which are either due to violating SQL specification, or missing or unclear SQL specification. A comparison with state-of-the-art RDBMS testing approaches shows that \tool{} detects more bugs and inconsistencies than those baselines during the same time period, and most of the bugs and inconsistencies cannot be detected by those baselines. We have made \tool{} public available to inspire further research. 
\section*{Acknowledgement}
This work is supported by the National Natural Science Foundation of China under Grant Nos. 62472429, 61972403, 61732014, 62102283), by Ant Group through CCF-Ant Research Fund No. CCF-AFSG RF20240103 and by the Ministry of Education, Singapore under its Academic Research Fund Tier 2 (Award ID: T2EP20222-0037).

\balance
\bibliographystyle{IEEEtranS}
\bibliography{ref}


\end{document}